%% file: main.tex
\documentclass{vldb}

\usepackage{graphicx}
\usepackage{balance}  

\usepackage[acronym]{glossaries}

\usepackage{tikz}
\usepackage{pgfplots}
\usepackage[caption=false]{subfig}
\usetikzlibrary{patterns}

\usepackage{multirow}
\usepackage{hhline}

\usepackage{textcomp}
\usepackage{color}
\usepackage{dblfloatfix}

\newcommand{\TODO}[1]{}
\newcommand{\parparaw}[1]{\texttt{ParPaRaw}}
\newcommand{\changed}[1]{#1}
\newcommand{\optswitch}[1]{}
\newcommand{\Description}[1]{}

\definecolor{grapefruit}{RGB}{255,143,128}
\definecolor{bavariablue}{RGB}{131,187,229}
\definecolor{grassgreen}{RGB}{163,217,119}
\definecolor{captpurple}{RGB}{222,95,133}
\definecolor{caliorange}{RGB}{255,77,0}

\definecolor{lightergreen}{HTML}{a3d977}
\definecolor{darkergreen}{HTML}{4E8F00}
\definecolor{darkerblue}{HTML}{0065BD}
\definecolor{davisblue}{HTML}{DAAA00}
\definecolor{nevadagold}{HTML}{FFC30F}

\definecolor{palblue}{RGB}{39, 50, 93}
\definecolor{palgold}{RGB}{245, 166, 1}
\definecolor{palorange}{RGB}{231, 66, 33}

\definecolor{nyccol}{HTML}{3CAEA3}
\definecolor{yelpcol}{RGB}{39, 50, 93}

\usepackage{url}

\usepackage{amsmath}

\usepackage[shortcuts]{extdash}

\usepackage{enumitem}
\pgfplotsset{compat=1.14}

\makeatletter
\newcommand\resetstackedplots{
\makeatletter
\pgfplots@stacked@isfirstplottrue
\makeatother
\addplot [forget plot,draw=none] coordinates{(0,0) (1,0) (2,0) (3,0)  (4,0)};
}
\makeatother

\vldbTitle{ParPaRaw: Massively Parallel Parsing of Delimiter-Separated Raw Data}
\vldbAuthors{Elias Stehle, and Hans-Arno Jacobsen}
\vldbDOI{https://doi.org/10.14778/3377369.3377372}
\vldbVolume{13}
\vldbNumber{5}
\vldbYear{2020}

\definecolor{grapefruit}{RGB}{255,143,128}
\definecolor{bavariablue}{RGB}{131,187,229}
\definecolor{grassgreen}{RGB}{163,217,119}
\definecolor{captpurple}{RGB}{222,95,133}

\input{pre_glossary.tex}

\begin{document}
\title{ParPaRaw: Massively Parallel Parsing of Delimiter-Separated Raw Data}

\numberofauthors{2} 

\author{
\alignauthor
Elias Stehle\\
  \affaddr{Technical University of Munich (TUM)}\\
  \email{stehle@in.tum.de}
\alignauthor
Hans-Arno Jacobsen\\
  \affaddr{Technical University of Munich (TUM)}\\
  \email{jacobsen@in.tum.de}
}

\maketitle

\begin{abstract}
\input{chap_abstract.tex}
\end{abstract}

\input{chap_introduction}
\input{chap_background}
\input{chap_related_work}
\input{chap_body}

\input{chap_evaluation}
\input{chap_conclusions}

\input{chap_appendix}
\input{chap_acknowledgements}

\bibliographystyle{abbrv}
\bibliography{bibliography}

\end{document}

%% file: pre_glossary.tex
\newacronym{csv}{CSV}{comma-separated values}
\newacronym{iot}{IoT}{Internet of Things}
\newacronym{gpu}{GPU}{Graphics Processing Unit}
\newacronym{cpu}{CPU}{Central Processing Unit}
\newacronym{simd}{SIMD}{single instruction, multiple data}
\newacronym{dfa}{DFA}{deterministic finite automaton}
\newacronym{pcie}{PCIe}{Peripheral Component Interconnect Express}
\newacronym{blob}{BLOB}{Binary Large Object}
\newacronym{utf}{UTF}{Unicode Transformation Format}
\newacronym{css}{CSS}{concatenated symbol string}
\newacronym{dbms}{DBMS}{Database Management System}
\newacronym{json}{JSON}{JavaScript Object Notation}
\newacronym{swar}{SWAR}{SIMD within a register}
\newacronym{fsm}{FSM}{finite state machine}
\newacronym{jit}{JIT}{just-in-time}
\newacronym{bfi}{BFI}{bit-field insert}
\newacronym{bfe}{BFE}{bit-field extract}

%% file: chap_abstract.tex
Parsing is essential for a wide range of use cases, such as stream processing, bulk loading, and in-situ querying of raw data.
Yet, the compute-intense step often constitutes a major bottleneck in the data ingestion pipeline,
since parsing of inputs that require more involved parsing rules is challenging to parallelise.
This work proposes a massively parallel algorithm for parsing delimiter-separated data formats on GPUs.
Other than the state-of-the-art, the proposed approach does not require an initial sequential pass over the input to determine a thread's parsing context.
That is, how a thread, beginning somewhere in the middle of the input, should interpret a certain symbol (e.g., whether to interpret a comma as a delimiter or as part of a larger string enclosed in double-quotes).
Instead of tailoring the approach to a single format, we are able to perform a massively parallel \gls{fsm} simulation,
which is more flexible and powerful, supporting more expressive parsing rules with general applicability.
Achieving a parsing rate of as much as 14.2 GB/s, our experimental evaluation on a GPU with $3 \thinspace 584$ cores shows that the presented approach is able to scale to thousands of cores and beyond.
With an end-to-end streaming approach, we are able to exploit the full-duplex capabilities of the PCIe bus and hide latency from data transfers.
Considering the end-to-end performance, the algorithm parses $4.8$ GB in as little as $0.44$ seconds, including data transfers.

%% file: chap_introduction.tex

\newcommand{\clf}{Common Log Format}
\newcommand{\elf}{Extended Log Format}

\section{Introduction}
Massive amounts of data from a wide range of data sources are made available using delimiter-separated formats,
such as \gls{csv}, and various log file formats like the \clf{} and the \elf{} \cite{2005:CSV,1995:CLF,1996:ELF}.
The relevancy of the \gls{csv} format, for instance, is highlighted by the plethora of public datasets, some in excess of hundreds of gigabytes in size, that are provided using the \gls{csv} format \cite{2016:taxi, 2018:kaggle}.
Log files are another origin of data in a delimiter-separated format that constitute an important source for many analytical workloads.
For instance, \emph{Sumo Logic}, a cloud-based log management and analytics service, recently announced that it analyses more than $100$ petabytes of data and 500 trillion records daily \cite{2018:sumo}.
With an ever increasing amount of data, there is also a growing need to provide and maintain rapid access to data in delimiter-separated formats.
This is also emphasised by ongoing research on in-situ processing of raw data and similar efforts that aim to lower the time to insight
\cite{
2019:fishstore,
2018:sparser,
2017:recache,
2016:constance,
2016:hetdata,
2015:vertpartraw,
2014:parscidata,
2014:speculative,
2014:raw,
2013:datavaults,
2012:nodb,
2011:idreos}.

While systems face an ever increasing amount of data that needs to be ingested and analysed, processors are seeing only moderate improvements in sequential processing performance.
In order to continue the trend of providing exponentially growing computational throughput, manufacturers have therefore progressively turned towards scaling the number of cores as well as extending \gls{simd} capabilities.
\glspl{gpu}, which have focused on parallelism ever since, now integrate as much as $5\thinspace120$ cores on a single chip \cite{2017:nvidiavolta}.
Further, CPUs comprising multiple \emph{chiplets}, as well as research focusing on package-level integration of multiple GPU modules, give an indication that hardware parallelism moves even beyond a single chip,
scaling to multiple inherently parallel \emph{chiplets} and GPU modules, respectively, on a package \cite{2017:MMG}.

In order to leverage the current degree of hardware parallelism and benefit from the ongoing trend of an ever growing number of cores,
algorithms have to be designed for massive scalability from the ground up \cite{2013:Johnson}.
Parsing, as a fundamental and compute-intense step in the data ingestion pipeline is no exception to this.
\optswitch{Independent of whether this is stream processing, in-situ processing of raw data, or bulk loading, parsing is essential for a wide range of use cases.}

\begin{sloppypar}
Parallel parsing of non-trivial delimiter-separated data formats, however, poses a great challenge,
as symbols have to be interpreted differently, depending on the context they appear in.
For the CSV format, for instance, RFC 4180 specifies that delimiters (i.e., commas and line breaks), which appear within a field that is enclosed in double-quotes,
have to be interpreted as part of the field, instead of being interpreted as actual field or record delimiters \cite{2005:CSV}.
In addition, many formats use a symbol to indicate comments or directives (e.g., \texttt{'\#'}),
following which, all symbols until the end of line have to be interpreted differently, yet again.
Since the context depends on all symbols preceding the symbol currently being interpreted,
it is impossible for a thread to simply begin parsing somewhere in the middle of the input.
Hence, the input cannot simply be split into multiple chunks that are processed independently.
\end{sloppypar}

\begin{figure}[!t]
\centering{
\includegraphics[width=1.0\linewidth]{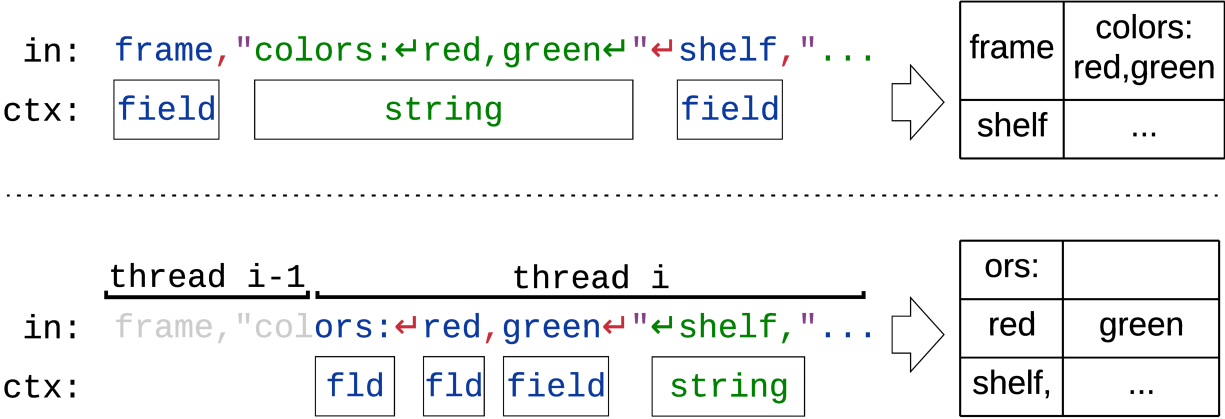}
}
\vspace{-0.5\baselineskip}
\caption{Challenges for parallel parsing: lacking context leads to misinterpretation}
\vspace{-0.9\baselineskip}
\label{fig:contextex}
\end{figure}

This is exemplified in Figure~\ref{fig:contextex}, where \emph{thread i} begins parsing in the middle of the input.
The thread is not aware of the double-quote preceding its chunk that indicates the beginning of a larger string, changing the parsing context.
As a result, the thread misinterprets subsequent commas and line breaks as delimiters, while they were actually supposed to be considered as part of the field's string.
A similar challenge arises for determining the records and columns that a chunk of the input belongs to,
which, again, depends on all the input preceding the current chunk being interpreted.
Finally, threads have to coordinate and possibly collaborate in order to assemble field values that span multiple threads. This may also involve converting symbols to a binary type (e.g., \texttt{int}, \texttt{float}).

Previous work on parallel loading of delimiter-separated data formats has addressed the challenge of determining a thread's parsing context by either performing an initial sequential pass over the input
or by completely dropping support for inputs with different parsing contexts, such as inputs containing enclosing symbols (e.g., double-quotes), comments, or directives \cite{2013:instant,2012:alenka}.
Another alternative is to tailor the approach to one specific format and exploit the format-specific characteristics, which, however, limits the approach's flexibility and applicability \cite{2017:mison,2019:langdalelemire,2018:simantex,2018:rapids,2019:speculative}.
One such exploit for a simple \gls{csv} format, for instance, is to count the number of double-quotes, inferring the beginning and end of enclosed strings depending on whether the count is odd or even, respectively.
\changed{
More recently, Ge et al. presented an approach for distributed \gls{csv} parsing \cite{2019:speculative}.
They aim to circumvent an initial sequential pass by exploiting \gls{csv}-specific characteristics to speculate about the parsing context.
While such a tailored approach works well with \gls{csv} as long as it strictly complies with the format expected by the algorithm,
it requires designing a completely new approach from the ground up once the input format deviates.
Parsing other delimiter-separated formats, such as log files and their multifaceted formats, poses a challenge for an approach that is tailored to \glspl{csv}.
Another important characteristic of state-of-the-art approaches is that they are designed for coarse-grained parallelism of distributed and multi-core systems,
which renders them infeasible for the fine-grained parallelism required by \glspl{gpu} \cite{2013:instant,2019:speculative}.
}

While constraining the input limits the applicability and flexibility,
performing a sequential pass over the input contributes a substantial portion of sequential work that limits scalability and, following Amdahl's law, precludes any speed-ups beyond a certain point.
Given the ongoing trend of increasing hardware parallelism on the one hand and the diversity of data sources that today's OLAP systems are confronted with on the other hand,
addressing these shortcomings is a viable endeavour.

%
We present \parparaw{}, an algorithm for massively \textbf{par}allel \textbf{pa}rsing of delimiter-separated \textbf{raw} data on \glspl{gpu} that overcomes these scalability issues without compromising applicability or constraining supported input formats.
\parparaw{} is designed from the ground up to scale linearly with the number of cores, providing robust performance despite the huge diversity of inputs it is confronted with,
by employing a data parallel approach with fine-grained parallelism.
\changed{\parparaw{} is designed to leverage the specifics of \glspl{gpu}.
It enables parallelism even beyond the granularity of a single record and ensures load balancing by splitting the input into small chunks of equal size that threads can process independently.}
Since using a data parallel approach raises the aforementioned challenges, we present an efficient solution for correctly identifying the parsing context of a thread's chunk, as well as its records and columns.
In order to provide a flexible approach that is applicable to a wide range of inputs, we allow specifying the parsing rules in the form of a \gls{dfa}.
\optswitch{We present a solution to deal with variable-width symbols comprising multiple bytes, which may cause symbols to cross chunk boundaries and span multiple chunks.
In addition, we address multiple challenges arising when targeting the fine-grained parallelism of \glspl{gpu},
due to their limited scratchpad memory, lack of efficient global synchronisation methods, and thread registers that do not support dynamic addressing.}
In order to exploit the full-duplex capabilities of the \gls{pcie} bus and lower the end-to-end latency, we present a streaming approach,
which parses data on the \gls{gpu}, while simultaneously transferring raw data to, and parsed data from the \gls{gpu}.

With a generally applicable approach that does not impose constraints on the input, we are able to parse as much as $14.2$~GB/s on the \gls{gpu}.
For end-to-end workloads, including data transfers via the \gls{pcie} bus,
\parparaw{} parses $4.8$~GB from the yelp reviews dataset in as little as $0.44$~seconds.
\optswitch{With a total processing time of only $0.91$ seconds, a similar performance is achieved for parsing $9.1$~GB of the nyc taxi trips dataset.}

\noindent
In summary, the contributions of this paper are four-fold.
\begin{enumerate}[leftmargin=*]
    \item We present an approach to massively parallel parsing of delimiter-separated data formats that is designed for scalability without sacrificing applicability and flexibility.
    The approach develops a scalable, data parallel algorithm that addresses three challenges:
    a) determining a thread's parsing context without requiring a prior sequential pass,
    b) determining the records and columns that a thread's symbols belong to, and
    c) efficiently coordinating threads to collaboratively generate field values.
    \vspace{-0.6\baselineskip}
    \item We address the major challenges that arise when mapping our algorithm to \glspl{gpu}, which provide only very limited addressable on-chip memory (tens of KB) and, due to their limited register file size, require lightweight threads with only very limited context.
    \vspace{-0.6\baselineskip}
    \item We show how to exploit the full-duplex capabilities of the \gls{pcie} bus with a streaming extension. This lowers the end-to-end latency and allows parsing data on the \gls{gpu}, while simultaneously transferring raw data to, and returning parsed data from the \gls{gpu}.
    \vspace{-0.6\baselineskip}
    \item Our experimental evaluation highlights that, given today's level of hardware parallelism, it is worth to design algorithms for scalability from the ground up, even if it implies a significant increase in the overall work being performed.
\end{enumerate}

This paper is organised as follows.
Section~\ref{sec:relatedwork} gives an overview of related approaches.
Section~\ref{sec:algorithm} presents the algorithm, its building blocks, and the processing steps.
Section~\ref{sec:optimisation} introduces optimisations, extensions, and implementation details.
Section~\ref{sec:evaluation} evaluates the presented approach. 

%% file: chap_background.tex

%% file: chap_related_work.tex
\section{Related Work}\label{sec:relatedwork}

Even though parsing is fundamental for in-situ processing of raw files and constitutes a major bottleneck in the data ingestion pipeline,
there is only limited work on accelerating the process.
This is also highlighted by Dziedzic et al., who show that modern \glspl{dbms} are unable to saturate available I/O bandwidth \cite{2016:loadingsurvey}.
Using a variety of hardware configurations and datasets, Dziedzic et al. provide an extensive analysis of the data loading process for multiple state-of-the-art \glspl{dbms} \cite{2016:loadingsurvey}.
Their evaluation reveals that data loading is CPU-bound \cite{2016:loadingsurvey}.

A notable advancement for parsing delimiter-separated formats is made by M{\"u}hlbauer et al. who present improvements along two lines \cite{2013:instant}.
On the one hand, they introduce optimisations to reduce the number of control flow branches by utilising \gls{simd} instructions for the identification of delimiters.
On the other hand, they present an approach for parallel parsing.
Their approach splits the input into multiple chunks of equal size that are processed in parallel.
Threads start parsing their chunk only from an actual record boundary onward, i.e., the first record delimiter in their chunk.
Threads continue parsing beyond their chunk until encountering the end of their last record.
This ensures that threads always process complete records, yet makes the approach sensible to the chosen chunk size and the input's record sizes.
For instance, the majority of threads, which work on a record that spans multiple chunks, unsuccessfully search for the beginning of their first record,
without performing actual parsing work.
Another shortcoming is that threads are not aware of the actual parsing context of their chunk.
That is, whether to interpret a field or record delimiter as an actual delimiter or as part of a field's value.
To address this, they introduce a \emph{safe mode} for formats that may contain more involved parsing rules.
In \emph{safe mode} a sequential pass over the input is performed, which keeps track of the parsing context, such as quotation scopes,
splitting chunks only at actual record delimiters.
\emph{Safe mode}, however, introduces a considerable portion of serial work, which, according to Amdahl's law,
precludes any speedup beyond a certain point.
By exploiting CSV-specific characteristics, Ge et al., who look at distributed CSV parsing, are able to bypass that initial pass by speculating about the parsing context \cite{2019:speculative}.
Similar to the approach of M{\"u}hlbauer et al., Ge et al. require coarse-grained parallelism (distributed parsing), as threads begin parsing only from a chunk's first record boundary onward.
As a result their approach is also sensible to the chosen chunk size and the input's record sizes.
The authors highlight that the performance of their approach degrades with decreasing chunk sizes, once a chunk approaches the size of a record \cite{2019:speculative}.
Moreover, both solutions do not provide parallelism beyond the granularity of an individual record, which makes them susceptible to load-balancing issues, particularly for small chunks and large, varying record sizes.
These circumstances render the two approaches infeasible for the fine-grained parallelism required by \glspl{gpu} \cite{2013:instant,2019:speculative}.

\glsunset{json}
\glsunset{jit}
Apart from work addressing delimiter-separated formats, multiple approaches tailored to processing \gls{json} have been proposed.
Li et al. present \emph{Mison}, a \gls{json} parser that supports projection and filter pushdown
by speculatively predicting logical locations of queried fields based on previously seen patterns \cite{2017:mison}.
Mison deviates from the classic approach of using an \gls{fsm} while parsing, which allows it to use \gls{simd} vectorisation to identify structural characters,
such as double-quotes, braces, and colons.
Whenever a structural character is encountered, its occurrence is recorded in the respective bitmap index (e.g., the double-quotes bitmap index).
The beginning and end of a string enclosed in double-quotes can be inferred from looking at the odd and even number of set bits, respectively.
While this enables \gls{simd} vectorisation and avoids branch divergence, circumventing the use of an \gls{fsm} and, hence, tailoring the approach specifically to the \gls{json} format,
limits the approach's applicability to formats with more involved parsing rules.
Bonetta et al. introduce \emph{FAD.js}, a runtime system for processing JSON objects that is based on speculative \gls{jit} compilation and selective access to data \cite{2017:fadjs}.
Palkar et al. propose a technique referred to as \emph{raw filtering}, which is applied on the data's raw bytestream before parsing \cite{2018:sparser}.
Langdale et al. recently introduced \emph{simdjson}, a standard-compliant, highly-efficient \gls{json} parser that makes use of \gls{simd} instructions.
Similar to \emph{Mison}, they focus on the \gls{json} format, which allows them to avoid the use of an \gls{fsm} while parsing \cite{2019:langdalelemire}.

The parallel prefix scan is a fundamental algorithm of \parparaw{} and a frequently recurring building block for data parallel algorithms.
Over the years many approaches for a parallel prefix scan have been proposed
\cite{2016:scan,2013:streamscan,2013:scalablegpuscan,2009:merrill,2008:gpuscan,1989:blelloch,1986:hillis,1982:brentkung,1973:koggestone,1960:sklansky}.
For a given binary reduction operator (e.g., addition), it takes an array of input elements and returns an array,
where the $i$-th output element is computed by applying the reduction operator to all input elements up to and including the $i$-th element \cite{2016:scan}:
$y_{i}=\bigoplus _{k=0}^{i}x_{k}$

A prefix scan that excludes the $i$-th input element is called exclusive prefix scan.
The prefix scan using addition is called prefix sum.
The following table shows an example of the inclusive and exclusive prefix sum, respectively:

\begin{center}
\scriptsize
\begin{tabular}{|r|c|c|c|c|c|c|c|c|}
\hline
$x_{i}$ & 3 & 5 & 1 & 2 & 9 & 7 & 4 & 2 \\
\hline\hline
$y_{i}$ (incl.) & 3 & 8 & 9 & 11 & 20 & 27 & 31 & 33 \\
\hline\hline
$y_{i}$ (excl.) & 0 & 3 & 8 & 9 & 11 & 20 & 27 & 31 \\
\hline
\end{tabular}
\end{center}

It is worth noting that all efficient parallel approaches require the binary operator to be associative.
The prefix scan used in \parparaw{} builds on the more recent work from Merrill et al., who propose a single-pass prefix scan \cite{2016:scan}.
Using the parallel prefix scan has also been considered for parallel parsing of regular languages.
In fact, the theory for parallel parsing dates back as long as four decades.
In particular, Fischer presents an algorithm that instantiates one \gls{fsm} for each state defined by the \gls{fsm} \cite{1975:fischer}.
Hillis et al. illustrate the use of the parallel prefix scan computation by presenting an algorithm that is similar to earlier work from Fischer \cite{1986:hillis}.
Even though the theory dates back several decades, it became feasible only more recently with modern hardware, i.e., GPUs with thousands of cores, that would set off the cost of running multiple \gls{fsm} instances.
We believe that this is the main reason why, to the best of our knowledge, this idea has not been considered for identifying the parsing context when parsing delimiter-separated formats on GPUs.
Our approach reconsiders the idea behind the work from Fischer and Hillis et al. to address the sub-problem of identifying the parsing context.
For identifying the parsing context, we devise a solution that respects the characteristics of \glspl{gpu}.
We show that, given todays degree of hardware parallelism, the cost of running multiple \gls{fsm} instances can be set off with our efficient and scalable solution.


%% file: chap_body.tex

\begin{figure*}[ht]
\centering{
\includegraphics[width=\textwidth]{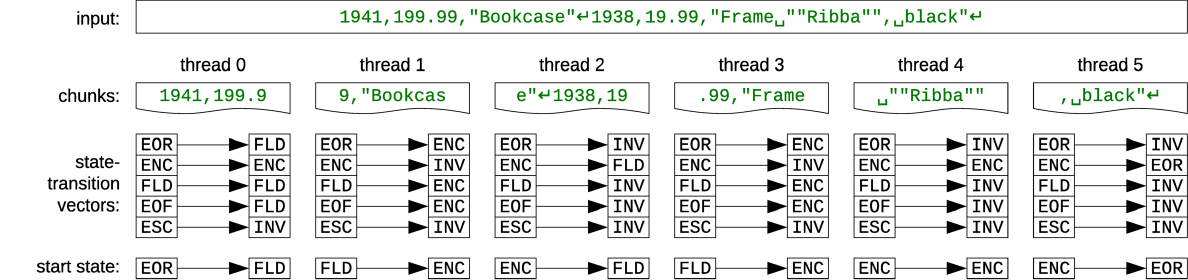}
}
\vspace{-1.2\baselineskip}
\caption{Determining the parsing context}

\label{fig:parser_one}
\end{figure*}
\section{Massively Parallel Parsing}\label{sec:algorithm}
In order to achieve scalability, even beyond thousands of cores, we pursue a data parallel approach, which splits the input into multiple chunks (e.g., $32$ bytes per chunk) that can be processed independently by the threads.
While a data parallel approach allows for massive scalability, there are three key challenges to overcome:

\begin{enumerate}[leftmargin=*]
    \vspace{-0.5\baselineskip}
    \item Determining the parsing context of a thread's chunk. That is, how a thread is supposed to interpret the symbols within its chunk (Section~\ref{ssec:parsing}).
    \vspace{-0.5\baselineskip}
    \item Determining the records and columns that the symbols of a thread's chunk belong to (Section~\ref{ssec:reccol}).
    \vspace{-0.5\baselineskip}
    \item Efficient coordination and collaboration between threads to transform a sequence of symbols to the data type of the respective column, e.g., \texttt{float}  (Section~\ref{ssec:partitioning}).
\end{enumerate}

\begin{sloppypar}
\parparaw{} addresses these challenges in multiple steps.
With each step, \parparaw{} gains additional information about each thread's chunk. This information is captured in meta data that subsequent steps can build on.
\end{sloppypar}

\subsection{Parsing}\label{ssec:parsing}

The first step addresses the challenge of identifying the parsing context of a thread's chunk, allowing a thread to meaningfully interpret its symbols.
That is, distinguishing whether a symbol is a control symbol (e.g., delimiting a field or a record) or whether it is part of a field's value.

It is important to note that without constraining the supported input formats and therefore sacrificing the approach's applicability, it is impossible to determine a thread's parsing context without considering all symbols preceding its chunk.
However, if a thread is supposed to consider all symbols preceding its chunk, the approach has to either perform an initial sequential pass over the input or wait for all threads working on preceding chunks to finish.
Considering all symbols preceding a thread's chunk introduces severe implications on the approach's scalability.

\parparaw{}, however, aims to neither constrain the input nor to introduce sequential work.
In order to achieve this, we exploit the fact that there are only few different contexts to consider while parsing.
While this increases the overall effort by a constant factor, it enables a fully concurrent approach and allows to scale linearly with the number of cores.

\begin{figure}[!b]
\centering
\includegraphics[width=0.75\linewidth]{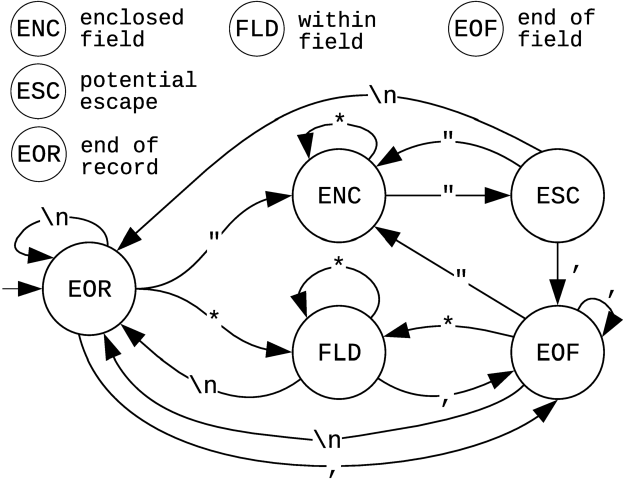}
\vspace{-0.6\baselineskip}
\caption{Example for a simple DFA parsing CSVs}
\vspace{-0.3\baselineskip}
\label{fig:parser_sm}
\end{figure}

In pursuit of a flexible approach that is generally applicable, \parparaw{} uses a \gls{dfa} while parsing.
The current parsing context is represented by the \gls{dfa}'s state. While a thread iterates over its symbols, it transitions the states of its \gls{dfa} according to its transition table.
One example of a \gls{dfa} for parsing a simple CSV format is shown in Figure~\ref{fig:parser_sm} (for simplicity it omits the invalid state (\texttt{INV}) used to track invalid formats, e.g., reading quotes in \texttt{FLD} state). 
A sequential approach would simply set the starting state of its \gls{dfa} and read the symbols of the input beginning to end, always being aware of the current state when reading a symbol.
For a data parallel approach, however, a thread, starting to parse somewhere in the middle of the input, cannot simply infer the state it is supposed to start in.

\begin{figure*}[ht]
\centering{
\includegraphics[width=0.99\textwidth]{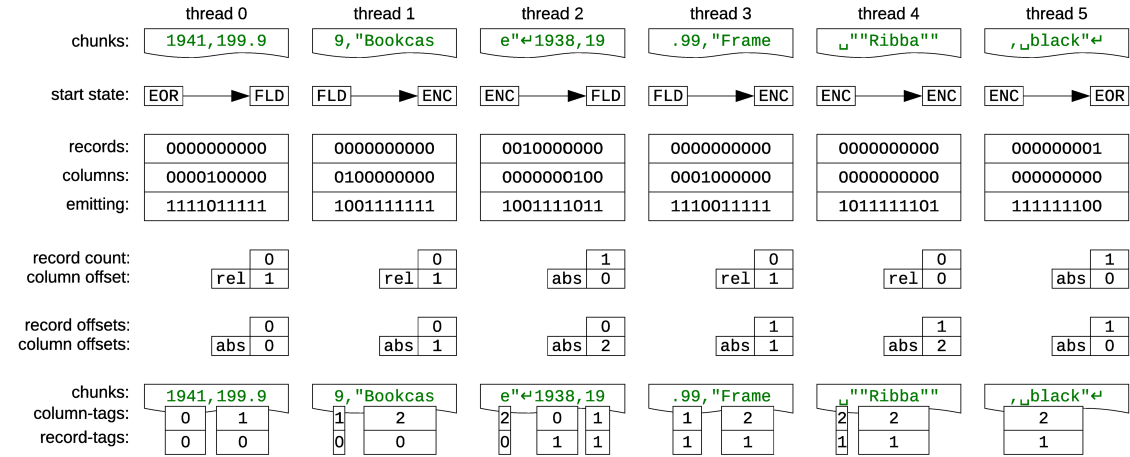}
}
\vspace{-0.9\baselineskip}
\caption{Identifying columns and records}
\vspace{-0.8\baselineskip}
\label{fig:parser_two}
\end{figure*}

\newcommand{\cofftsub}{t}
\newcommand{\coffsub}{{o}}
\newcommand{\matdeli}{\\}
\newcommand{\cpt}{CPT}

In order to perform meaningful work despite lacking the correct starting state, each thread instantiates one \gls{dfa} for every state, $s_{i} \in S$, defined by the \gls{dfa}, setting the starting state of the $i$-th \gls{dfa}-instance to state $s_{i}$ (see \emph{state-transition vectors} in Figure~\ref{fig:parser_one}).
For reasons of clarity, in the following we assume $s_{i}=i$, i.e., representing a state by its index, to avoid the intricate differentiation between a state and a state index.
An efficient implementation uses the same mechanism during preprocessing to ensure efficient lookups into data structures like the transition table.
While the thread is reading the symbols of its chunk, it transitions the states of all its \gls{dfa}-instances accordingly.
Once all the symbols of a chunk have been read, the final state of each \gls{dfa}-instance is noted in a state-transition vector.
We maintain one state-transition vector per thread, with each state-transition vector holding $|S|$ elements.
The $i$-th entry of the state-transition vector represents the final state of the $i$-th \gls{dfa}-instance (i.e., the \gls{dfa}-instance that has originally started in state $s_{i}$).
Hence, the algorithm can infer that if a thread had started parsing in state $s_{i}$, it would end up in the state given by the $i$-th entry of that thread's state-transition vector after the thread has read all its symbols (see Figure~\ref{fig:parser_one}).

By computing the composite of these state-transition vectors, the algorithm can deduce the starting state for every thread.
We define the composite operation $a \circ b$ of two state-transition vectors $a$ and $b$ as:
\vspace{-0.2\baselineskip}
\newcommand{\cmstdeli}{\\}
\begin{equation*}
\left[{\begin{smallmatrix}a_{0} \cmstdeli a_{1} \cmstdeli \vphantom{\int\limits^x}\smash{\vdots} \cmstdeli a_{|S|-1}\end{smallmatrix}}\right] \circ \left[{\begin{smallmatrix}b_{0} \cmstdeli b_{1} \cmstdeli \vphantom{\int\limits^x}\smash{\vdots} \cmstdeli b_{|S|-1}\end{smallmatrix}}\right]  =
\left[{\begin{smallmatrix}b_{a_{0}} \cmstdeli b_{a_{1}} \cmstdeli \vphantom{\int\limits^x}\smash{\vdots} \cmstdeli b_{a_{|S|-1}}\end{smallmatrix}}\right]
\end{equation*}
\vspace{-0.5\baselineskip}

Since the composite operation is associative, the algorithm can compute a parallel exclusive scan using the composite operation, which is seeded with the identity vector.
After the exclusive scan, the $i$-th entry of each thread's resulting vector now corresponds to the state that the thread's \gls{dfa} is supposed to start in,
if the sequential \gls{dfa}'s starting state was $s_{i}$.
For instance, if the sequential \gls{dfa}'s starting state was $s_{3}$, each thread finds its starting state by reading the element at index three from its resulting vector.

Once each thread is aware of its starting state, threads can correctly interpret the symbols from their chunk by simulating a single \gls{dfa}-instance.
While iterating over its symbols, a thread identifies field delimiters, record delimiters, and other control symbols (e.g., an escape sequence)
according to the parsing rules specific to the current format being parsed.
Since the algorithm addresses delimiter-separated formats, the relevant meta data for each symbol can be represented using three bitmap indexes:
one marking symbols that are delimiting a record,
one flagging symbols that are delimiting a field,
and one indicating whether a symbol is a control symbol (e.g., escape symbol) or whether it is part of the field.
Subsequent steps can build on these bitmap indexes without requiring to repeatedly simulate the \gls{dfa}-instance.

\begin{figure*}[!t]
\centering{
\includegraphics[width=\textwidth]{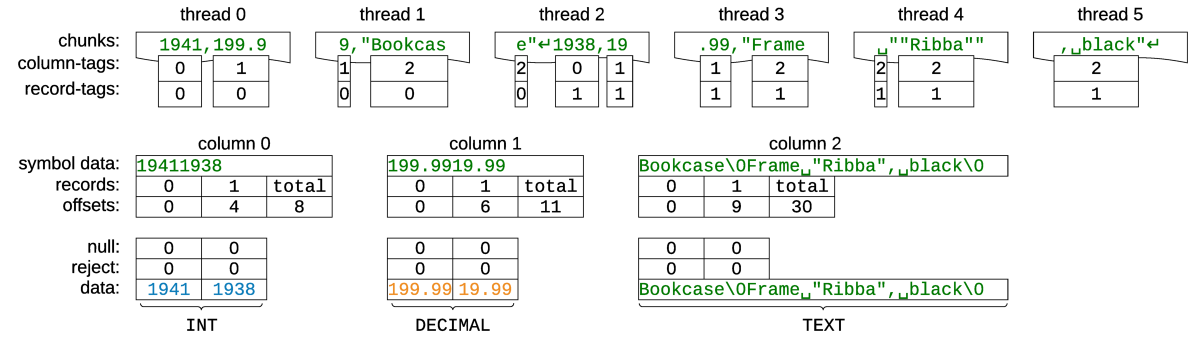}
}
\caption{Preparing data in a columnar format and type conversion}
\label{fig:parser_three}
\end{figure*}

\subsection{Identifying Columns and Records}\label{ssec:reccol}
The bitmap indexes from the previous step are used to identify the column and record offset.
That is, the record and column that the first few symbols of a thread's chunk belong to, until it encounters the first delimiting symbol.
Determining the column and record offsets requires two steps.

First, each thread computes the offset that its chunk adds to the preceding chunk's offset.
For the records, the relative offset can easily be computed by counting the records (i.e., the number of set bits of a thread's record delimiter bitmap index using \texttt{POPCNT}).
For columns, however, this is slightly more involved.
If a thread encounters a record delimiter and therefore the beginning of a new record, it can infer the \textbf{absolute} column offset for the subsequent chunk (e.g., \emph{thread~2} in Figure~\ref{fig:parser_two}).
Otherwise, all it can infer is that it has seen $k$ field delimiters and, therefore, the next chunk's column offset has an additional offset of $k$, \textbf{relative} to the preceding chunk's column offset.
In Figure~\ref{fig:parser_two}, for instance, \emph{thread~1} encounters one column delimiter but no record delimiter.
As \emph{thread~1} is not yet aware of its own column offset, it can only infer that the subsequent thread's column offset increases by one, relative to its own column offset.
We distinguish between an absolute and a relative column offset, which are denoted as \texttt{abs} and \texttt{rel}, respectively, in Figure~\ref{fig:parser_two}.
A column offset is absolute, if there is at least one set bit in the thread's record delimiter bitmap index.
The column offset can be computed by zeroing all bits of the column delimiter bitmap index that precede the last set bit in the record delimiter bitmap index, counting the remaining set bits, i.e.:
\texttt{ POPCNT( ({\raise.17ex\hbox{$\scriptstyle\mathtt{\sim}$}}BLSMSK(rec\_bidx)) \& col\_bidx )}

In a subsequent step, the algorithm computes the exclusive prefix sum over the record counts, which yields each thread's record offset.
In order to retrieve the column offsets, we perform an exclusive scan using the following operation, where,
for a column offset $x$, $x_{\cofftsub}$ denotes whether a column offset is relative (\texttt{rel}) or absolute (\texttt{abs}) and $x_{\coffsub}$ denotes the offset value:

\begin{equation*}
\left[{\begin{smallmatrix}a_{\cofftsub} \matdeli a_{\coffsub}\end{smallmatrix}}\right] \oplus \left[{\begin{smallmatrix}b_{\cofftsub} \matdeli b_{\coffsub}\end{smallmatrix}}\right] =
\begin{cases}
\left[{\begin{smallmatrix}b_{\cofftsub} \matdeli b_{\coffsub}\end{smallmatrix}}\right] & \text{if } b_{\cofftsub} \text{ is }\texttt{abs}\\[10pt]
\left[{\begin{smallmatrix}a_{\cofftsub} \matdeli a_{\coffsub} + b_{\coffsub}\end{smallmatrix}}\right] & \text{if } b_{\cofftsub} \text{ is }\texttt{rel} \\
\end{cases}
\end{equation*}

Once all absolute column and record offsets have been calculated, threads can correctly identify the column and record that each of its symbols belongs to.
In preparation for the next step, which transforms the row-oriented input to a columnar format and, if applicable, converts strings of symbols to the data type of the corresponding column,
we tag the symbols with the column and record they belong to, as illustrated at the bottom of Figure~\ref{fig:parser_two}.

\subsection{Columnar Format \& Type Conversion}\label{ssec:partitioning}
Now, that each thread is fully aware of the associated columns and records, threads still need to generate the individual field values in a columnar format.
Depending on the column that a string of symbols belongs to, this may require converting to the respective column's type (e.g., \texttt{int}, \texttt{float}).
As shown at the top of Figure~\ref{fig:parser_three}, symbols belonging to the same field may still span multiple chunks, requiring involved threads to collaborate on generating a single field value.
To circumvent the collaboration between threads entirely, one option would be to change the assignment of threads, assigning exactly one thread to exclusively process all symbols required for generating a single field value.
This approach, however, may cause considerable load-balancing issues, as the number of symbols per field may be subject to high variance.
In particular, values of columns with variable-width types, such as text or \glspl{blob}, may be arbitrarily large.

Another challenge arises due to the fact that symbols are still in a row-oriented format.
That is, two threads working on subsequent chunks may be parsing two different columns of two different types.
Hence, they may require different rules, executing completely different code segments for parsing the fields' values.
For instance, one thread may require generating an integer value, while the next one is extracting a date.
The behaviour of different threads executing different code paths is particularly punishing on \glspl{gpu},
where all threads within a warp (e.g., a group of $32$ threads) are executing the same instruction in lockstep.

\parparaw{} addresses these challenges by first partitioning all symbols by the column they are associated with.
During partitioning, \parparaw{} ensures that symbols within a column maintain their order by using a stable radix sort that uses the symbols' column-tags as the sort-key.
While sorting, the symbols and the record-tags are moved along with the associated sort-key.
The radix sort iterates over the bits of the column-tags, performing a stable partitioning pass on the sequence of bits considered with a given pass.
A single partitioning pass involves (1) computing the histogram over the number of items that belong to each partition,
(2) computing the exclusive prefix sum over the histogram's counts, and (3) scattering the items to the respective partition.

After partitioning, all symbols belonging to the same column lie cohesively in memory.
We refer to all symbols belonging to the same column as the \gls{css} of a column.
The histogram that is maintained while sorting is used to identify the offsets of the columns' \glspl{css}.
Similar to the symbols, all the symbols' record-tags lie cohesively in memory, indicating which record a symbol belongs to.

Having all symbols in a columnar format allows the algorithm to efficiently process each of the columns.
This may include type inference, validation, identifying \texttt{NULL}s, and converting symbol strings to the column's type.
First, \parparaw{} uses the record-tags to generate an index into the \gls{css}.
The index is used to identify the offsets and lengths of the fields' symbol strings.
To generate the index, the algorithm performs a run-length encoding on the symbols' record-tags, which yields each field's record and its number of symbols.
Computing the exclusive prefix sum over the fields' symbol counts yields the offsets into the \gls{css}, as shown in Figure~\ref{fig:parser_three}.
The symbol count of a field can be inferred using the difference of the successive field's offset and the field's own offset.

Building on the index, \parparaw{} can now start generating the fields' values by interpreting the strings of symbols, if that is required for a given column (e.g., numerical or temporal types).
In order to address possible load-balancing issues due to having high variance in the number of symbols per field,
we use three different collaboration levels: thread-exclusive, block-level, and device-level collaboration.
By default, a thread tries to exclusively generate a field value, looking up the offset and number of its symbols in the index.
Once the thread has identified the symbols, it starts converting the symbol string to the column's type (e.g., \texttt{int}, \texttt{float}).
If, during lookup, a thread detects that its string of symbols exceeds a certain threshold, it will defer generating that field value for the block- or device-level collaboration.
The threshold depends on the on-chip memory of a \gls{gpu}'s streaming multiprocessor and its number of cores.
If there are fields left for the block-level collaboration, all threads of a thread-block (e.g., $64$ threads) collaborate on generating a field value.
Fields that exceed the on-chip memory available to a thread-block (typically in the order of tens of kilobytes) are addressed by the device-level collaboration.
Block- and device-level collaboration use the same data parallel approach as the overall approach presented for parsing delimiter-separated inputs.
Hence, the same technique for determining a thread's parsing context is employed.

\section{Extensions \& Implementation}\label{sec:optimisation}

Having presented the fundamental processing steps for a robust approach to massively parallel parsing in Section~\ref{sec:algorithm},
this section focuses on optimisations, extensions, and implementation details.
We develop two optimised specialisations that can be applied if a given input meets certain conditions (see Section~\ref{ssec:fastmode}).
Section~\ref{ssec:varlen} addresses symbols crossing chunk boundaries, such as being encountered when dealing with variable-length encodings.
To highlight that not only efficiency but also the approach's applicability was of great importance to this work,
we present a few more capabilities in Section~\ref{ssec:capabilities}.
With an end-to-end streaming extension, we aim to hide the latency of data transfers via the \gls{pcie} bus (see Section~\ref{ssec:streaming}).
Finally, Section~\ref{sec:implementation} presents how we address the major challenges of mapping the algorithm to the \gls{gpu}.

\subsection{Alternative Tagging Modes}\label{ssec:fastmode}
\parparaw{}, as presented in Section~\ref{sec:algorithm}, focuses on robustness.
It is even resilient to inputs that contain records with a varying number of field delimiters per record (e.g., "\texttt{1,Apples\textbackslash{}n2\textbackslash{}n}").
This section focuses on presenting two optimised specialisations that are chosen,
if the input provides a constant number of columns per record or if the user prefers to reject records that have an inconsistent number of field delimiters.

\begin{figure}[!t]
\centering{
\includegraphics[width=0.9\linewidth]{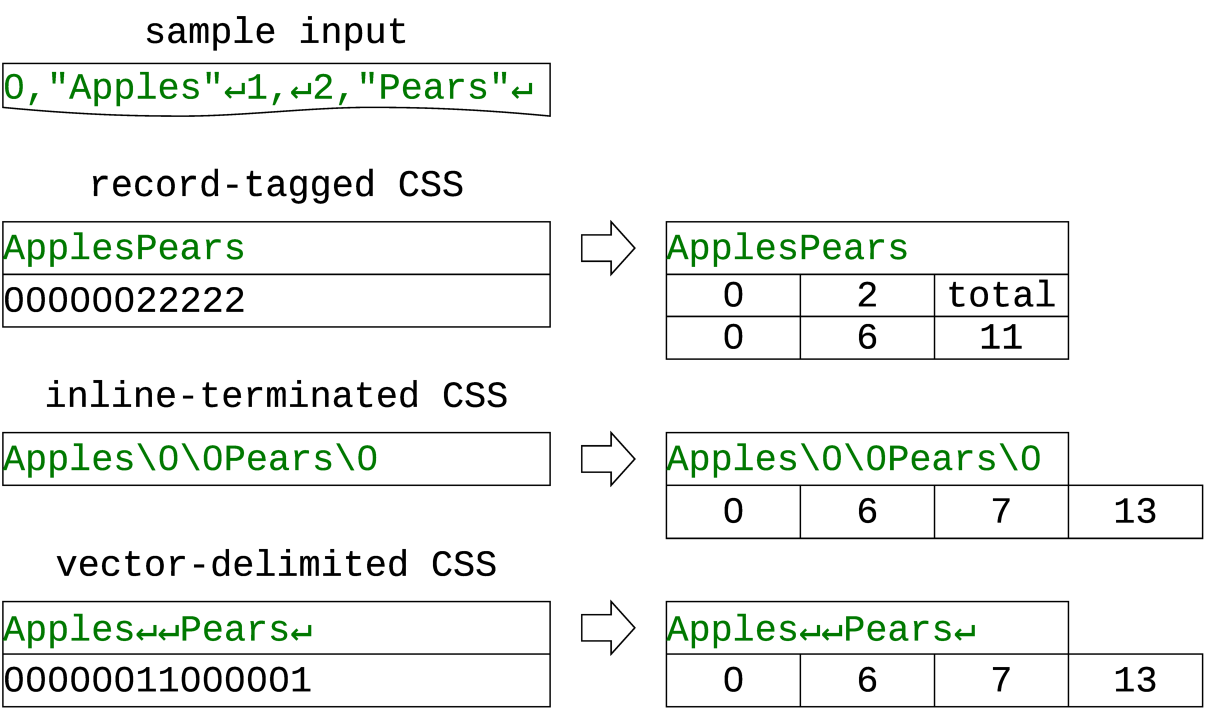}
}
\vspace{-0.3\baselineskip}
\caption{Alternative tagging modes}
\vspace{-0.9\baselineskip}
\label{fig:alt_modes}
\end{figure}

Since many of the presented processing steps work at peak memory bandwidth, reading and writing record-tags of four bytes increases the amount of memory transfers and degrades performance.
Hence, we aim to lower the amount of memory transfers by reducing the memory footprint of the record-tags.
As illustrated in Figure~\ref{fig:alt_modes}, we provide two alternatives to record-tags.

The \emph{inline-terminated \gls{css}} replaces delimiters with a terminator during the tagging phase.
Just like the null character for null-terminated strings, the terminator is a unique character that indicates the end of a field's symbols.
Good candidates for terminators are various separators specified by the \texttt{ASCII} standard, such as the \emph{record separator} (\texttt{0x1E}) or the  \emph{unit separator} (\texttt{0x1F}).
To generate the \gls{css}'s index, the algorithm simply writes the offsets of all occurrences of the terminator symbols to the index.
The \emph{inline-terminated \gls{css}} requires that the terminator is not part of a column's \gls{css},
as those symbols would otherwise get confused for a terminator.

The \emph{vector-delimited \gls{css}} can address this scenario by devoting its own auxiliary boolean vector that delimits the fields within a column.
The \gls{css}'s index is generated the same way as for the \emph{inline-terminated \gls{css}}
with the minor difference that the algorithm identifies non-zero values in the auxiliary vector instead of terminators from the \gls{css}.

\subsection{Variable-Length Symbols}\label{ssec:varlen}

So far, we have not addressed the challenge of symbols crossing chunk boundaries.
While this can be easily prevented for fixed-size symbols spanning multiple bytes by adjusting the chunk size to be an integer multiple of the symbol size,
it is more involved for variable-length symbols.
For instance, if inputs are encoded using a variable-length \gls{utf}, such as \gls{utf}\=/8 or \gls{utf}\=/16,
symbol boundaries become unpredictable and some symbols might be crossing chunks.
If a symbol crosses chunk boundaries, the thread working on the chunk at which the symbol begins (i.e., the symbol's leading bytes) is in charge of reading that symbol and transitioning the state of its \gls{dfa} accordingly.
Threads working on subsequent chunks that only read trailing bytes of a symbol skip those bytes.
For the variable-length encodings \gls{utf}\=/8 and \gls{utf}\=/16, threads can identify whether the first bytes of a chunk are only trailing bytes of an encoded code point (a code point is a numerical value and most code points are assigned a character).
\gls{utf}\=/8 encodes code points using one, two, three, or four bytes.
Unless a single byte is used, all trailing bytes have the common binary prefix of \texttt{0b10XX}\thinspace{}\texttt{XXXX}.
Hence, for \gls{utf}\=/8 encoded inputs, threads simply ignore a chunk's first few bytes with that binary prefix.
\gls{utf}\=/16 uses either two bytes to encode code points ranging from \texttt{0x0000} to \texttt{0xD7FF} and from \texttt{0xE000} to \texttt{0xFFFF},
and four bytes for code points beyond \texttt{0x010000}.
If four bytes are used, the two high order bytes, referred to as high surrogate, are in the range of \texttt{0xD800} to \texttt{0xD8FF},
and the low order bytes, referred to as low surrogate, are in the range of \texttt{0xDC00} to \texttt{0xDFFF}.
Since unicode does not assign any characters in the range of \texttt{0xD800} to \texttt{0xDFFF},
there is no two-byte combination in that range.
Hence, similar to \gls{utf}\=/8, a thread ignores a chunk's first two bytes if their value is in the range of \texttt{0xDC00} to \texttt{0xDFFF}.

\subsection{Capabilities}\label{ssec:capabilities}

This section focuses on pointing out a few more capabilities to highlight \parparaw{}'s applicability to real-world requirements.

\textbf{Validating format} --- One notable strength of \parparaw{} is its ability to simulate an \gls{fsm} while parsing,
which makes it widely applicable and enables more expressive parsing rules.
With the presented massively parallel approach for simulating a \gls{dfa}, \parparaw{} is always aware of the \gls{dfa}'s current state when reading a symbol.
Hence, invalid state transitions as well as a non-accepting end state can easily be detected.

\textbf{Skipping records and selecting columns} --- \parparaw{} is able to ignore a user-specified set of records and columns.
While tagging symbols with their associated column and record, all symbols that belong to records or columns that are supposed to be ignored are identified and marked as irrelevant.
Irrelevant symbols can be ignored following the partitioning step.

\textbf{Skipping rows} -- It is worth noting that rows are different from records, as some records may span multiple rows.
Since ignoring rows may interfere with the assignment of symbols to columns and records, \parparaw{} has to ensure that rows are ignored early on.
Hence, \parparaw{} ignores a set of rows by performing an initial parallel pass over the input, pruning symbols of ignored rows (i.e., parallel stream compaction).

\textbf{Inferring or validating number of columns} --- If no schema is provided and therefore the number of columns is not known a priori, \parparaw{} can infer the input's number of columns.
Similarly, if \parparaw{} is supposed to reject records that do not conform to the expected number of columns the same technique is applied.
In either case, during \gls{dfa} simulation threads need to track three values in addition to the relative or absolute column offset handed over to the subsequent chunk.
Firstly, every thread keeps track of the number of field delimiters encountered before reading its very first record delimiter, which subsequently is referred to as \emph{relative min/max}.
Further, every thread maintains the minimum and maximum number of columns it counted per record for all records following the chunk's first record delimiter.
We use an extra bit to denote if no minimum and maximum was determined, i.e., the chunk does not contain any record delimiter.
After the prefix scan of the column offsets, \parparaw{} can resolve the \emph{relative min/max}, turning it into an absolute column offset.
The absolute column offset is then incorporated in the respective chunk's minimum and maximum column count.
A subsequent reduction over the maximum is then used to infer the number of columns.
Comparing the identified minimum and maximum column counts indicates whether a given chunk conforms to the expected number of columns per record.

\textbf{Default values for empty strings} --- If the input has a consistent number of field-delimiters per record,
the default value for empty strings is set during type conversion.
That is, when field values are parsed, the empty string is parsed as the column's default value.
If the input does not have a consistent number of field-delimiters per record,
the column's data is pre-initialised with the user-specified default value and later overwritten for non-empty fields.

\begin{sloppypar}
\textbf{Type inference} --- \parparaw{} is comparably efficient when identifying a column's type, as, prior to type conversion, all of a column's symbols lie cohesively in memory.
During an initial pass over the column's symbols, threads identify the minimum numerical type being required to back their field value.
A subsequent parallel reduction over the minimum type yields the inferred type of a column.
\parparaw{} currently only considers type inference for numerical types, but can be extended to cover temporal types.
\end{sloppypar}

\subsection{End-to-End Streaming}\label{ssec:streaming}

This section provides an extension to \parparaw{}'s on-\gls{gpu} parsing algorithm presented in Section~\ref{sec:algorithm}
to address inputs that do not reside on the \gls{gpu} or exceed its available device memory.
In order for the \gls{gpu} to be able to process the input, the input first needs to be transferred via the comparably slow \gls{pcie} bus
and, once processed, the parsed data has to be returned.
It is worth noting that the \gls{pcie} bus allows for full-duplex communication, enabling simultaneous data transfers in either direction at peak bandwidth.
While the \gls{pcie} bus does not necessarily limit the throughput, waiting for the data transfer to complete before and after parsing, respectively,
adds a considerable amount of latency to the end-to-end processing time.
Hence, rather than waiting for the input to arrive on the \gls{gpu}, before the \gls{gpu} begins processing it and, once finished, starts returning the parsed data,
we make use of a streaming approach.
The streaming approach splits the input into multiple partitions.
Each partition, at some point, is transferred to the \gls{gpu}, processed, and its data is returned.
Having multiple partitions allows to overlap these stages for subsequent partitions, similar to the pipelined approach in \cite{2016:shahvarani, 2017:stehle}.
That is, transferring a partition, while processing its predecessor on the \gls{gpu} and simultaneously returning parsed data via the interconnect.

\begin{figure}[t]
\centering{
\includegraphics[width=1.0\linewidth]{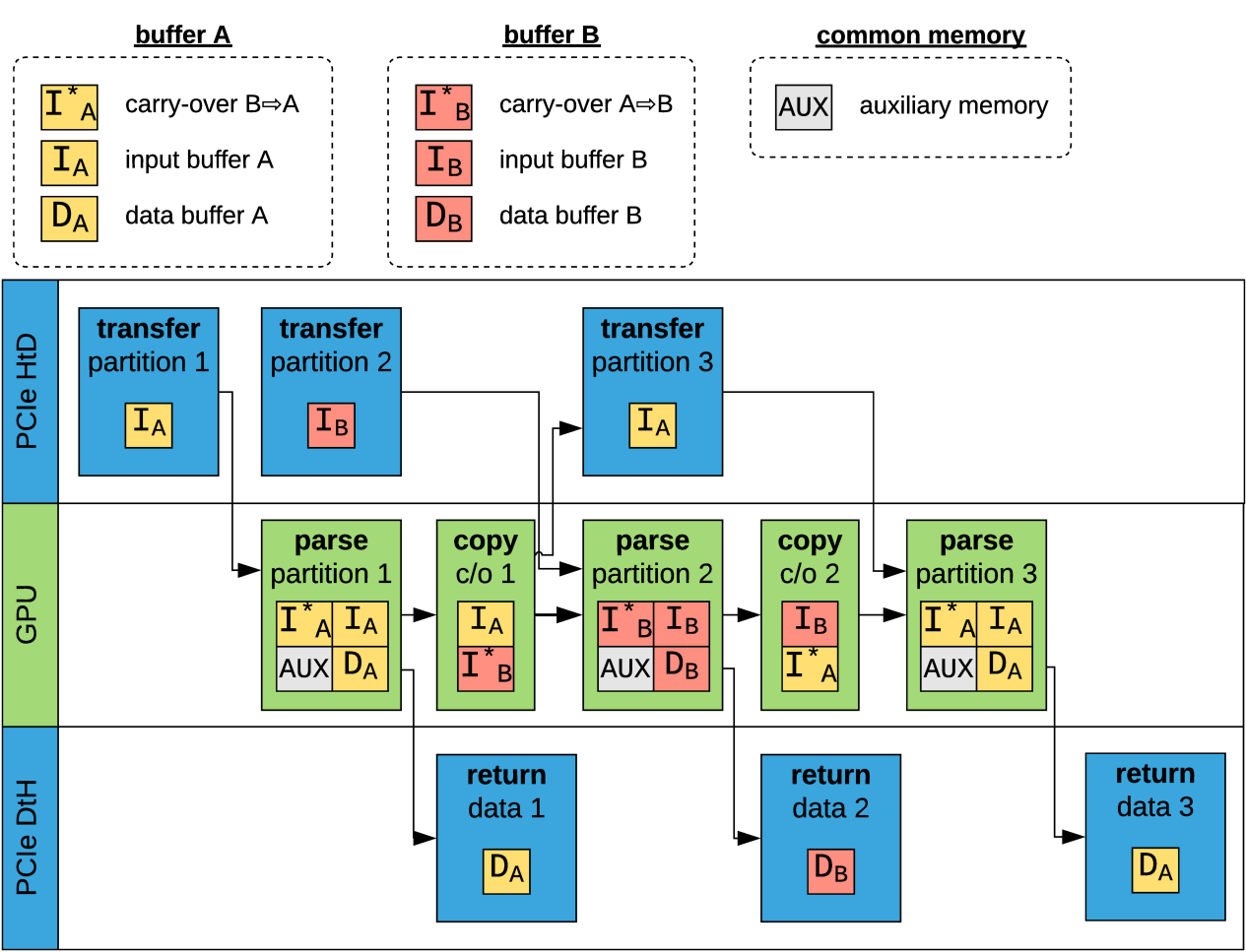}
}
\vspace{-1.2\baselineskip}
\caption{End-to-end streaming}
\vspace{-0.9\baselineskip}
\label{fig:pipelined}
\end{figure}

\begin{figure*}[t]
    \begin{minipage}[c]{\textwidth}
      \centering
      \includegraphics[width=\textwidth]{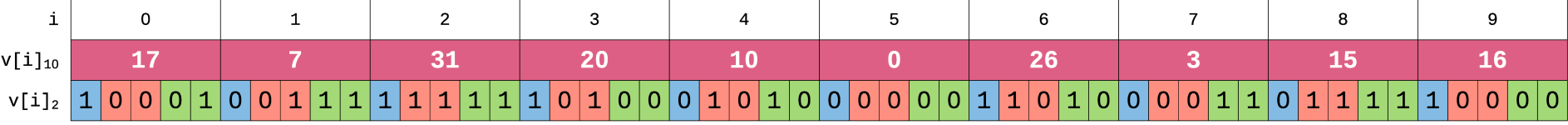}
      \vspace{-0.2cm}
    \end{minipage}
  \begin{minipage}[c]{.69\textwidth}
    \centering
    \includegraphics[width=\textwidth]{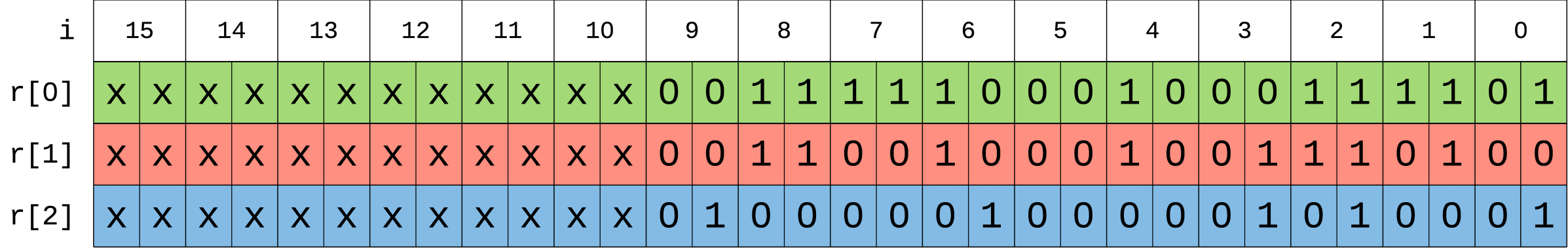}
  \end{minipage}\hfill
\bgroup
\def\arraystretch{1.25}
\begin{minipage}{.30\textwidth}
    \centering
    \small
    \begin{tabular}{|l|c|c|}
    \hline
    num. items                    & c & 10 \\ \hline
    bits per item                 & b & 5  \\ \hline
    \begin{tabular}[c]{@{}l@{}}avail. bits per\vspace{-2pt} \\ item-fragment\end{tabular} & $\text{a}=\left\lfloor \frac{32}{\text{c}}\right\rfloor$ & 3  \\ \hline
    \begin{tabular}[c]{@{}l@{}}bits per\vspace{-2pt} \\ item-fragment\end{tabular}        & $\text{k}=2^{\left\lfloor\log_2 \text{a} \right\rfloor}$ & 2  \\ \hline
        fragments                     & $\left\lceil \text{b}/\text{k}\right\rceil$ & 3  \\ \hline
    \end{tabular}
\end{minipage}
\egroup
\vspace{-0.3\baselineskip}
\caption{Logical and physical view of the multi-fragment in-register array}
\vspace{-0.2\baselineskip}
\label{fig:multifrag}
\end{figure*}

\begin{sloppypar}
For the end-to-end streaming approach, we allocate a double-buffer and some auxiliary memory on the \gls{gpu} (see top of Figure~\ref{fig:pipelined}).
Each buffer comprises memory for the raw input and the parsed data.
One buffer's raw input allocation is used as input for parsing on the \gls{gpu},
while the opposing buffer's raw input allocation is receiving data of the next partition.
Similarly, one buffer's data allocation is used to output parsed data,
while data is being returned via the interconnect from the opposing buffer's data allocation.
In addition\optswitch{ to allocating sufficient memory to hold the input for one partition},
we prepend additional memory for a carry-over to the memory allocated for the input of each buffer.
The carry-over is used for prepending the last, incomplete record at the end of one buffer's input to the opposing buffer's input.
\end{sloppypar}

Figure~\ref{fig:pipelined} exemplifies the processing steps of the streaming parsing approach.
The stages of a partition are
(1) \emph{transfer}: transferring the raw input of a partition from the host to the \gls{gpu},
(2) \emph{parse}: parsing the input of a given partition, including the prepended carry-over and writing the parsed data to the data buffer, and
(3) \emph{return}: returning the parsed data from the data buffer to the host.
The resources required by each processing step are illustrated by the rectangular symbols within a step (e.g., $I_{A}$ representing the memory allocated for the input of buffer $A$).
A processing step's dependency on a preceding processing step is depicted by an incoming edge.
An important sequence depicted in Figure~\ref{fig:pipelined} is when the \gls{gpu} switches work from one double-buffer to the other.
For instance, after the \gls{gpu} has finished parsing the input of the first partition (raw input provided by \emph{input buffer A}),
the last incomplete record is prepended to the second partition by copying it to the memory of the \emph{carry-over} of \emph{buffer B}.
Since copying the carry-over is reading from \emph{input buffer A},
the algorithm ensures that the transfer of the third partition to \emph{input buffer A} does not take place before the carry-over has been copied, as the carry-over would otherwise get corrupted.

\subsection{Implementation Details}\label{sec:implementation}

This section addresses the main challenges faced when mapping \parparaw{} to the \gls{gpu}.
Specifically, we introduce a new data structure, referred to as multi-fragment in-register array (MFIRA),
which provides a workaround for the constraint that threads cannot dynamically address into the register file.
Since the register file is extremely fast and provides the most on-chip memory, addressing this shortcoming is a viable endeavour.
The presented data structure allows to dynamically index into and access elements of a bounded array.
MFIRA is particularly efficient for low-cardinality arrays of small integers.
This is a recurring pattern in \gls{gpu} programming, since the \gls{gpu}'s threads need to be very lightweight, allowing for only very limited context (i.e., using only few registers).
Hence, even though MFIRA was designed as an efficient data structure backing various objects when parsing,
MFIRA likely would be useful for other use cases as well.
Further, we present a branchless algorithm that builds on \gls{swar} to identify the index of a read symbol in the transition table.
With that approach, we are able to keep the symbols that the algorithm compares against in the very fast register file.
At the same time, it avoids that threads within a warp are executing along different branches.

\textbf{Multi-fragment in-register array} ---
\optswitch{This data structure is fundamental for the approach's \gls{dfa} simulation.
For instance, it is used for the state-transition vector, when matching symbols,
and for the transition table, if it is small enough.
}
The idea behind the data structure is that even though thread registers themselves cannot be addressed dynamically,
individual bits within a register can be.
Specifically, we use the intrinsic functions \gls{bfi} and \gls{bfe}, which require only two clock cycles on recent microarchitectures,
to efficiently access an arbitrary sequence of bits from a register.
We use these two functions to decompose an item that is written to the data structure and distribute the item's fragments (i.e., partitions of its bits) amongst one or more registers.
Similarly, when an item is accessed, it is reassembled from its fragments.
Figure~\ref{fig:multifrag} illustrates this principle, depicting an array containing up to ten items, each five bits wide.
For such an array, the data structure could use up to three bits per fragment.
To efficiently compute bit-offsets into a register, however, the number of bits actually being used by the data structure is chosen to be a power of two.
This allows replacing the expensive integer multiplication with a bit-shift operation. 
In the example depicted in Figure~\ref{fig:multifrag}, the data structure would therefore devote two bits per fragment, using a total of three fragments.
The individual fragments of the items are colour-coded in Figure~\ref{fig:multifrag} to highlight how the logical view (top of the figure)
maps to the physical view (bottom of the figure).

\textbf{Symbol matching using \gls{swar}} ---
During \gls{dfa} simulation, the algorithm uses a transition table to identify the state transition from the \gls{dfa}'s current state and a read symbol to the \gls{dfa}'s new state.
The transition table is two-dimensional, with states along one and symbols along the other dimension.
To compress the transition table's size, we collapse all the transition table's symbols that have identical state transitions into symbol groups.
As illustrated in Table~\ref{tab:ttable}, we have one symbol group per row instead of having symbol groups as columns, which allows coalesced access to all state transitions of a read symbol.
This is particularly useful when computing the state-transition vectors.
A thread reads a symbol from its chunk, identifies its symbol group, and fetches the row of state transitions for the matched symbol group.
For each of its \gls{dfa} instances, it can then efficiently determine the new state from that row.

\bgroup
\def\arraystretch{1.2}
\begin{table}[t]
\caption{Transition table example}
\small
\begin{tabular}{|c|c|cccccc|}
\hline
\multirow{2}{*}{\begin{tiny}symbols\end{tiny}} & \multirow{2}{*}{\begin{tiny}groups\end{tiny}} & \multicolumn{6}{c|}{states}                                                                                                                \\ \cline{3-8}
                         &                         & \multicolumn{1}{c|}{EOR} & \multicolumn{1}{c|}{ENC} & \multicolumn{1}{c|}{FLD} & \multicolumn{1}{c|}{EOF} & \multicolumn{1}{c|}{ESC} & INV \\ \hline
\textbackslash{}n        & 0                       & EOR                      & ENC                      & EOR                      & EOR                      & EOR                      & INV \\
"                        & 1                       & ENC                      & ESC                      & INV                      & ENC                      & ENC                      & INV \\
,                        & 2                       & EOF                      & ENC                      & EOF                      & EOF                      & EOF                      & INV \\
*                        & 3                       & FLD                      & ENC                      & FLD                      & FLD                      & INV                      & INV \\ \hline
\end{tabular}
\label{tab:ttable}
\end{table}
\egroup

\begin{figure*}[!b]
\vspace{-0.5\baselineskip}
\captionsetup[subfigure]{labelformat=empty}
\subfloat[]{%
\begin{tikzpicture}
\begin{axis}[
    ybar stacked,
    reverse legend,
    legend cell align={left},
    height=0.77\columnwidth,
    width=0.5\textwidth,
    ybar=0pt,
    ymin=0,
    ymax=68,
    xlabel near ticks,
    ylabel near ticks,
    ylabel=processing duration (ms),
    xlabel style={align=center},
    xlabel={chunk size (in bytes)\\\vspace{-0.5\baselineskip}\\(a) yelp reviews},
    yticklabel={\pgfmathparse{\tick}\pgfmathprintnumber{\pgfmathresult} },
    y tick label style={/pgf/number format/fixed,
    /pgf/number format/1000 sep = \thinspace},
    nodes near coords align={vertical},
    xtick={0,12,28,44,60},
    xticklabels={4,16,32,48,64},
    enlarge y limits=0,
    bar width=2pt,
    minor y tick num=5,
    ymajorgrids,
    yminorgrids,
    tick align=inside,
    enlarge x limits=0.04,
    every axis plot/.append style={ultra thin},
    legend style={at={(0.5,0.98)},anchor=north, legend columns=3, transpose legend},
    xtick style={draw=none}
]

	\addplot[bar shift=0, fill=black!100, area legend] table[skip first n=0,x expr =  {\coordindex}, y expr=(\thisrowno{1}),col sep=comma,each nth point={1},comment chars={T}] {data/yelp_chunks.csv};
	\addplot[bar shift=0, fill=black!80, area legend] table[skip first n=0,x expr =  {\coordindex}, y expr=(\thisrowno{2}),col sep=comma,each nth point={1},comment chars={T}] {data/yelp_chunks.csv};
	\addplot[bar shift=0, fill=black!60, area legend] table[skip first n=0,x expr =  {\coordindex}, y expr=(\thisrowno{3}),col sep=comma,each nth point={1},comment chars={T}] {data/yelp_chunks.csv};
	\addplot[bar shift=0, fill=black!40, area legend] table[skip first n=0,x expr = {\coordindex}, y expr=(\thisrowno{4}),col sep=comma,each nth point={1},comment chars={T}] {data/yelp_chunks.csv};
	\addplot[bar shift=0, fill=black!20, area legend] table[skip first n=0,x expr =  {\coordindex}, y expr=(\thisrowno{5}),col sep=comma,each nth point={1},comment chars={T}] {data/yelp_chunks.csv};

    \legend{parse,scan,tag,partition,convert}
\end{axis}
\end{tikzpicture}
\label{fig:chunksizeyelp}
}
\hfill
\subfloat[]{%
\centering
\begin{tikzpicture}
\begin{axis}[
    ybar stacked,
    reverse legend,
    legend cell align={left},
    height=0.77\columnwidth,
    width=0.5\textwidth,
    ybar=0pt,
    ymin=0,
    ymax=68,
    xlabel near ticks,
    ylabel near ticks,
    ylabel=processing duration (ms),
    xlabel style={align=center},
    xlabel={chunk size (in bytes)\\\vspace{-0.5\baselineskip}\\(b) NYC taxi trips},
    yticklabel={\pgfmathparse{\tick}\pgfmathprintnumber{\pgfmathresult} },
    y tick label style={/pgf/number format/fixed,
    /pgf/number format/1000 sep = \thinspace},
    nodes near coords align={vertical},
    xtick={0,12,28,44,60},
    xticklabels={4,16,32,48,64},
    enlarge y limits=0,
    bar width=2pt,
    minor y tick num=5,
    ymajorgrids,
    yminorgrids,
    tick align=inside,
    enlarge x limits=0.04,
    every axis plot/.append style={ultra thin},
    legend style={at={(0.5,0.98)},anchor=north, legend columns=3, transpose legend},
    xtick style={draw=none}
]

	\addplot[bar shift=0, fill=black!100, area legend] table[skip first n=0,x expr =  {\coordindex}, y expr=(\thisrowno{1}),col sep=comma,each nth point={1},comment chars={T}] {data/nyc_chunks.csv};
	\addplot[bar shift=0, fill=black!80, area legend] table[skip first n=0,x expr =  {\coordindex}, y expr=(\thisrowno{2}),col sep=comma,each nth point={1},comment chars={T}] {data/nyc_chunks.csv};
	\addplot[bar shift=0, fill=black!60, area legend] table[skip first n=0,x expr =  {\coordindex}, y expr=(\thisrowno{3}),col sep=comma,each nth point={1},comment chars={T}] {data/nyc_chunks.csv};
	\addplot[bar shift=0, fill=black!40, area legend] table[skip first n=0,x expr = {\coordindex}, y expr=(\thisrowno{4}),col sep=comma,each nth point={1},comment chars={T}] {data/nyc_chunks.csv};
	\addplot[bar shift=0, fill=black!20, area legend] table[skip first n=0,x expr =  {\coordindex}, y expr=(\thisrowno{5}),col sep=comma,each nth point={1},comment chars={T}] {data/nyc_chunks.csv};

    \legend{parse,scan,tag,partition,convert}
\end{axis}
\end{tikzpicture}
\label{fig:chunksizenyc}
}
\vspace{-1.6\baselineskip}
\caption{Time spent on individual processing steps depending on the chunk size configuration}
\vspace{-1\baselineskip}
\label{fig:chunksize}
\end{figure*}
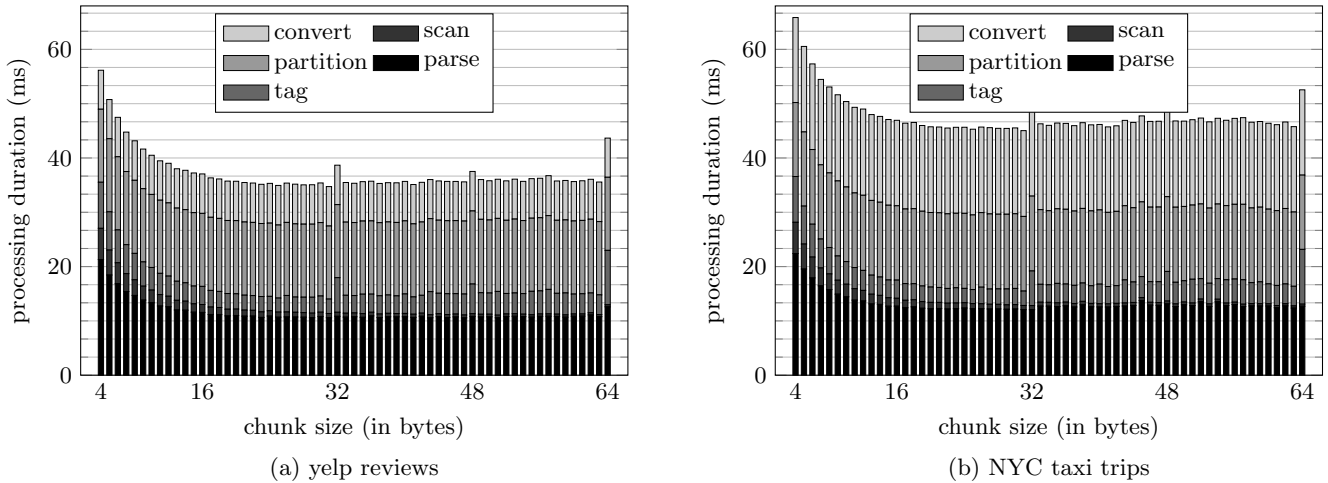

Having introduced symbol groups, mapping a symbol to its symbol group is an elementary step.
To ensure an efficient mapping, we exploit the fact that delimiter-separated formats typically have only a few symbols to distinguish amongst,
such as an escaping symbol, field and record delimiters, and enclosing symbols like quotes or brackets (see Table~\ref{tab:ttable}).
Hence, for the symbols we use a comparison-based approach, rather than devoting a full lookup-table that maps each character value to its group.
Since symbols are often only eight bits wide (e.g., \texttt{ASCII} and \gls{utf}\=/8-encoded \texttt{ASCII} characters), while \glspl{gpu} implement 32-bit wide arithmetic instructions,
we use a branchless \gls{swar} algorithm to perform multiple comparisons at a time (see Table~\ref{tab:twiddling}).
On the one hand, this avoids inefficiencies due to threads executing divergent branches.
On the other hand, with the following approach, we are more space-efficient and are able to keep the symbols in the very fast register file.
As illustrated in Table~\ref{tab:twiddling}, we place each of the symbols that we try to match against in the individual bytes of four-byte registers.
We refer to these registers as lookup-registers (LU-registers).
For later comparison against the LU-registers, whenever a symbol is read, we replicate that symbol in every byte of a separate register (i.e., the $s$-register).
Computing the exclusive or for each of the LU-registers with the $s$-register yields a null-byte if the two bytes match.
Subsequently applying the bit-twiddling hack to determine a null-byte, as suggested by Mycroft in 1987 \cite{1987:mycroft}, sets the most-significant bit for that byte (see definition of $H(x)$ in Table~\ref{tab:twiddling}).
Using the intrinsic function \texttt{bfind}, we retrieve the position of the most-significant set bit.
If no bit was set, i.e., the read symbol does not match any byte from the LU-registers, \texttt{bfind} will return \texttt{0xFFFFFFFF}.
To retrieve the matching index, we divide the value returned by \texttt{bfind} by eight, using bit-shift for efficiency reasons (i.e., shifting it three bits to the right).
For LU-registers that contain no match, the matching index is \texttt{0x1FFFFFFF}, while for the ones that contain a match, it yields a value between zero and three.
To ensure that we consider a match, if present, we compute the minimum over all matching indexes.
Finally, in case there was no match, we map the matching index of \texttt{0x1FFFFFFF} to the \texttt{catch-all} symbol group by using the minimum function.
The minimum is computed very efficiently, requiring only one or two cycles on recent microarchitectures and is therefore generally preferable to a conditional expression.

\begin{table}[t]
\caption{Identifying a symbol's index using SWAR}
\begin{tabular}{r|cccc||cccc|}
\hhline{~|----||----|}
byte   & 7 & 6 & 5 & 4 & 3 & 2 & 1 & 0 \\
\hhline{-|----||----|}
symbol group  &  &  & 3 & 2 & 2 & 2 & 1 & 0 \\
lookup ($LU$) &   &   &  & \texttt{\textbackslash{}t} & | & , & " & \texttt{\textbackslash{}n} \\
\hhline{-|----||----|}
read symbol ($s$) & ,  & , & , & , & , & , & , & , \\
$c=LU $ \texttt{XOR} $s$         & {\footnotesize\texttt{--}}    & {\footnotesize\texttt{--}}    & {\footnotesize\texttt{--}}    & {\footnotesize\texttt{25}}    & {\footnotesize\texttt{50}}    & {\footnotesize\texttt{00}}    & {\footnotesize\texttt{0E}}    & {\footnotesize\texttt{26}}    \\
$swar=H(c)$                         & {\footnotesize\texttt{--}}    & {\footnotesize\texttt{--}}    & {\footnotesize\texttt{--}}    & {\footnotesize\texttt{00}}    & {\footnotesize\texttt{00}}    & {\footnotesize\texttt{80}}    & {\footnotesize\texttt{00}}    & {\footnotesize\texttt{00}}    \\
$\texttt{bfind}(swar)\texttt{>>}3$                         & {\footnotesize\texttt{1F}}    & {\footnotesize\texttt{FF}}    & {\footnotesize\texttt{FF}}    & {\footnotesize\texttt{FF}}    & {\footnotesize\texttt{00}}    & {\footnotesize\texttt{00}}    & {\footnotesize\texttt{00}}    & {\footnotesize\texttt{02}}    \\
\hhline{-|----||----|}
\multicolumn{1}{r}{$idx=\min_{\forall r_i{}} (x)$} & \multicolumn{4}{c}{} & \multicolumn{4}{r}{\texttt{0x00000002}} \\
\multicolumn{1}{r}{$\min(idx, 5))$} & \multicolumn{4}{c}{} & \multicolumn{4}{r}{\texttt{\underline{0x00000002}}} \\
\hhline{---------}
\multicolumn{9}{r}{$H(x)=((x - \texttt{0x01010101})\ \&{}\ (\sim{}\! x)\ \&{}\ \texttt{0x80808080})$}
\end{tabular}
\vspace{-0.7\baselineskip}
\label{tab:twiddling}
\end{table}

%% file: chap_evaluation.tex
\section{Experimental Evaluation}\label{sec:evaluation}

\begin{sloppypar}
For the experimental evaluation we use two systems, one to evaluate CPU-only implementations, referred to as \emph{CPU system},
and one system equipped with a \gls{gpu} (\emph{GPU system}) used for evaluating \gls{gpu}-based approaches.
Both systems are running  Ubuntu 18.04.
The \emph{CPU system} has four sockets, each equipped with a Xeon E5-4650 clocked at 2.70 GHz.
It has a total of 512 GB of DRAM (DDR3-1600).
The \emph{GPU system} is equipped with 128 GB of DRAM (DDR4-2400) and a Xeon E5-1650 v4 processor with six physical cores, clocked at 3.60 GHz.
The source code was compiled with the \texttt{-O3} flag.
We used release 10.1.105 of the CUDA toolkit.
The \emph{GPU system} hosts an NVIDIA Titan X (Pascal) with 12 GB device memory, 3\thinspace584 cores, and a base clock of 1\thinspace417 MHz (driver version is 418.40.04).
\end{sloppypar}

The output of \parparaw{} is configured to comply with the format specified by \emph{Apache Arrow}.
\emph{Apache Arrow} specifies a columnar memory format for efficient analytic operations \cite{apachearrow}.
It is used by a multitude of well-known in-memory analytics projects, such as \emph{OmniSci}, \emph{pandas}, and \emph{Apache Spark}.
For \parparaw{} we use a \gls{dfa} that is capable of parsing any RFC4180 compliant input \cite{2005:CSV}.
The \gls{dfa} defines six states, including one state to track invalid state transitions.

To evaluate the systems, we choose the two dissimilar real-world datasets \emph{yelp reviews} and \emph{NYC taxi trips}.
The \emph{yelp reviews} dataset comprises $6.69$ million reviews from yelp's dataset as \gls{csv}, with all fields enclosed in double-quotes \cite{yelp}.
The dataset is $4.823$ GB large with an average record size of $721.4$ bytes per record.
Each record is made up of nine columns, covering text-based, numerical, and temporal types.
The dataset is of particular interest due to the text-based reviews that may include field and record delimiters,
which poses a challenge for many parallel parsers.

The \emph{NYC taxi trips} dataset is $9.073$ GB large and comprises $102.8$ million yellow taxi trips taken in the year 2018 provided by the \emph{NYC Taxi \& Limousine Commission} \cite{2016:taxi}.
The dataset's $17$ columns cover numerical and temporal datatypes.
With an average of only $88.3$ bytes per record and $5.2$ bytes per field,
the majority of the fields are very short and of a numerical type, putting the emphasis on data type conversion.

\subsection{On-GPU Parsing}\label{ssec:ongpu}

This section provides a detailed evaluation of the presented algorithm using on-GPU workloads.
Our on-GPU evaluation focuses on identifying efficient configurations and analysing the algorithm's sensibility to input parameters.
Time measurements for the on-\gls{gpu} parsing experiments represent the GPU wall-clock time, measured using CUDA events.
Other than the end-to-end parsing experiments, on-\gls{gpu} experiments do not include data transfers between host and device.
Unless noted otherwise, we use the first $512$ MB of each dataset for this evaluation, to be able to evaluate all tagging modes before running out of device memory.

We provide a breakdown of the time spent on the individual processing steps as a function of chunk size in Figure~\ref{fig:chunksize}.
Comparing the breakdown of the two datasets highlights the complexity of converting the many numerical and temporal types of the \emph{NYC taxi trips} dataset,
which, on average, make up only $5.2$ bytes per value.
The type conversion of the \emph{NYC taxi trips} dataset accounts for roughly one third of the total processing time.
Type conversion of the \emph{yelp reviews} dataset, in contrast, only contributes approximately $20\%$ to the total processing time,
as the text-based reviews make up the majority of the raw record size.
The analysis shows that the approach is mostly agnostic to choice of the chunk size, as long as it is reasonably large.
Only for tiny chunk sizes of $15$ bytes and less, the overhead of initialising and scheduling tens of millions of threads becomes noticeable.
For a tiny chunk size, the ratio of actual work being done in relation to the time spent on initialising threads and the amount of meta data being written becomes unfavourable.
Choosing a small chunk size is disadvantageous to parsing, tagging, and the prefix scan.
As the prefix scan's complexity is linear in the total number of chunks, its share of the processing time becomes noticeable when using very small chunks.
The prefix scan takes less than two percent of the total processing time for most choices of the chunk size.
The small spikes for parsing and tagging when using $32$, $48$, and $64$ bytes per chunk, respectively,
are due to shared-memory bank conflicts and bad occupancy.
The best performance is achieved for $31$ bytes per chunk, which will be used as default for the remaining evaluations.

\begin{figure}
\begin{tikzpicture}
\begin{axis}[
    height=0.7\columnwidth,
    width=\columnwidth,
    ybar=0.5pt,
    xlabel near ticks,
    ylabel near ticks,
    scaled y ticks = false,
    xlabel style={align=center},
    xlabel=input size (MB),
    ylabel=parsing rate (GB/s),
    axis x line*=bottom,
    yticklabel={\pgfmathparse{\tick}\pgfmathprintnumber{\pgfmathresult} },
    y tick label style={/pgf/number format/fixed,
    /pgf/number format/1000 sep = \thinspace},
    nodes near coords align={vertical},
    xtick={0,1,2,3,4,5,6,7,8,9,10},
    x tick label style={
        rotate=90,anchor=east,/pgf/number format/.cd,
        fixed,
        fixed zerofill,
        precision=2,
        /tikz/.cd
    },
    xticklabels={1,2,4,8,16,32,64,128,256,512},
    ymin=0,
    enlarge y limits={upper},
    bar width=5pt,
    minor y tick num=5,
    ymajorgrids,yminorgrids,
    ytick align=inside,
    enlarge x limits=0.1,
    legend style={at={(0.02,0.98)},anchor=north west}
]
	\addplot[fill=nyccol!100, draw=none, area legend] table[skip first n=0, x expr={\coordindex},y expr=\thisrowno{1}/1000000000*(1000/\thisrowno{2}),col sep=comma] {data/partition_nyc_sample.csv};
	\addplot[fill=yelpcol!100, draw=none, pattern color=white, area legend] table[skip first n=0, x expr={\coordindex},y expr=\thisrowno{1}/1000000000*(1000/\thisrowno{2}),col sep=comma] {data/partition_yelp_sample.csv};
	\legend{NYC taxi trips, yelp reviews}
\end{axis}
\end{tikzpicture}
\vspace{-0.8\baselineskip}
\caption{Parsing rate as a function of input size}
\vspace{-0.8\baselineskip}
\label{fig:inputsize}
\end{figure}
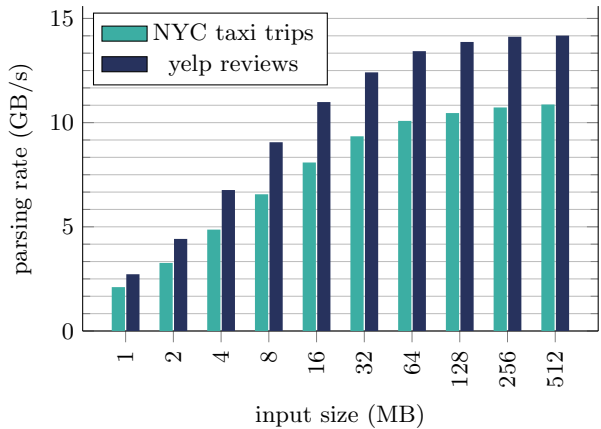

Figure~\ref{fig:inputsize} shows \parparaw{}'s performance for various different input sizes.
Parsing ten megabytes of the \emph{yelp reviews} dataset in as little as one millisecond, \parparaw{} shows impressive performance even for small inputs, achieving a parsing rate of $9.75$ GB/s.
For even smaller inputs, \parparaw{} is able to process a single megabyte from either dataset in less than $500$ $\mu{}s$,
corresponding to a parsing rate of more than $2.1$ GB/s and $2.7$ GB/s for the \emph{NYC taxi trips} and the \emph{yelp reviews} dataset, respectively.
Even though the absolute performance is impressive, in particular when compared to available parsers (see Section~\ref{ssec:endtoend}),
\parparaw{}'s efficiency degrades as the input size decreases.
When parsing only five megabytes of either of both datasets, \parparaw{}'s performance achieves roughly $50\%$ of its peak performance.
A major reason for this, especially for inputs that are parsed in less than a millisecond, is the overhead due to the many kernel invocations during the type conversion step.
During type conversion, there are multiple kernel invocations per column, required for the \gls{css}-index generation as well as the type conversion itself.
Hence, considering the many columns of the two datasets, kernel invocations, each with an estimated overhead in the order of roughly $5$~-~$10$ $\mu{}s$,
account for a reasonable share of the few hundred microseconds that are required for parsing those tiny inputs.

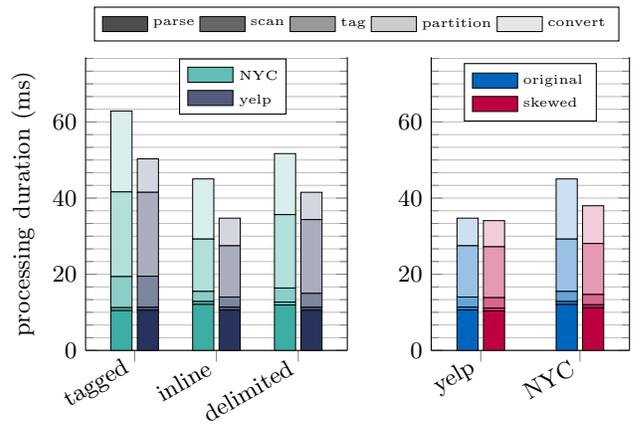
\begin{figure}
    \hspace{1.2cm}\ref{named}
\vspace{-0.5\baselineskip}
\\
\captionsetup[subfigure]{labelformat=empty}
\subfloat[]{%
\begin{tikzpicture}[baseline]
\begin{axis}[
    ybar stacked,
    legend cell align={left},
    height=0.65\columnwidth,
    width=0.6\columnwidth,
    ybar=0pt,
    xlabel near ticks,
    ylabel near ticks,
    scaled y ticks = false,
    xlabel style={align=center},
    xlabel=,
    ylabel=processing duration (ms),
    axis x line*=bottom,
    yticklabel={\pgfmathparse{\tick}\pgfmathprintnumber{\pgfmathresult} },
    y tick label style={/pgf/number format/fixed,
    /pgf/number format/1000 sep = \thinspace},
    nodes near coords align={vertical},
    xtick={0,1,2},
    legend style={font=\tiny},
    x tick label style={
        rotate=30,anchor=east,/pgf/number format/.cd,
        fixed,
        fixed zerofill,
        precision=2,
        /tikz/.cd
    },
    xmax=2,
    ymax=70,
    xticklabels={tagged,inline,delimited},
    ymin=0,
    enlarge y limits={upper},
    bar width=8pt,
    minor y tick num=5,
    ymajorgrids,yminorgrids,
    ytick align=inside,
    enlarge x limits=0.3,
    every axis plot/.append style={ultra thin},
    legend style={at={(0.65,0.98)},anchor=north},
    legend entries={parse,scan,tag,partition,convert},
    legend columns=5,
    legend to name=named
]

\addlegendimage{fill=black!70,area legend}
\addlegendimage{fill=black!60,area legend}
\addlegendimage{fill=black!40,area legend}
\addlegendimage{fill=black!20,area legend}
\addlegendimage{fill=black!10,area legend}
	\addplot[bar shift=5pt, fill=yelpcol!100, pattern color=white, area legend] table[skip first n=0,x expr={\coordindex},y expr=\thisrowno{1},col sep=comma,each nth point={1},comment chars={T}] {data/tags_yelp.csv};
	\addplot[bar shift=5pt, fill=yelpcol!80, pattern color=white, area legend] table[skip first n=0,x expr={\coordindex},y expr=\thisrowno{2},col sep=comma,each nth point={1},comment chars={T}] {data/tags_yelp.csv};
	\addplot[bar shift=5pt, fill=yelpcol!60, pattern color=white, area legend] table[skip first n=0,x expr={\coordindex},y expr=\thisrowno{3},col sep=comma,each nth point={1},comment chars={T}] {data/tags_yelp.csv};
	\addplot[bar shift=5pt, fill=yelpcol!40, pattern color=white, area legend] table[skip first n=0,x expr={\coordindex},y expr=\thisrowno{4},col sep=comma,each nth point={1},comment chars={T}] {data/tags_yelp.csv};
	\addplot[bar shift=5pt, fill=yelpcol!20, pattern color=white, area legend] table[skip first n=0,x expr={\coordindex},y expr=\thisrowno{5},col sep=comma,each nth point={1},comment chars={T}] {data/tags_yelp.csv};

\resetstackedplots
\addlegendimage{fill=yelpcol!80, pattern color=white,area legend}\label{p1}
\addlegendimage{black,fill=nyccol!80,area legend}\label{p2}

    \addplot[bar shift=-5pt, fill=nyccol!100, area legend] table[skip first n=0,x expr={\coordindex},y expr=\thisrowno{1},col sep=comma,each nth point={1},comment chars={T}] {data/tags_nyc.csv};
	\addplot[bar shift=-5pt, fill=nyccol!80, area legend] table[skip first n=0,x expr={\coordindex},y expr=\thisrowno{2},col sep=comma,each nth point={1},comment chars={T}] {data/tags_nyc.csv};
	\addplot[bar shift=-5pt, fill=nyccol!60, area legend] table[skip first n=0,x expr={\coordindex},y expr=\thisrowno{3},col sep=comma,each nth point={1},comment chars={T}] {data/tags_nyc.csv};
	\addplot[bar shift=-5pt, fill=nyccol!40, area legend] table[skip first n=0,x expr={\coordindex},y expr=\thisrowno{4},col sep=comma,each nth point={1},comment chars={T}] {data/tags_nyc.csv};
	\addplot[bar shift=-5pt, fill=nyccol!20, area legend] table[skip first n=0,x expr={\coordindex},y expr=\thisrowno{5},col sep=comma,each nth point={1},comment chars={T}] {data/tags_nyc.csv};
\end{axis}
\node [draw,fill=white] at (rel axis cs: 0.75,0.9) {\shortstack[l]{\tiny
\ref{p2} NYC \\\tiny
\ref{p1} yelp }};
\end{tikzpicture}
}
\hfill
\subfloat[]{%
\begin{tikzpicture}[baseline]
\begin{axis}[
    ybar stacked,
    reverse legend,
    legend cell align={left},
    height=0.65\columnwidth,
    width=0.5\columnwidth,
    ybar=0pt,
    xlabel near ticks,
    ylabel near ticks,
    scaled y ticks = false,
    xlabel style={align=center},
    xlabel=,
    ylabel=,
    axis x line*=bottom,
    yticklabel={\pgfmathparse{\tick}\pgfmathprintnumber{\pgfmathresult} },
    y tick label style={/pgf/number format/fixed,
    /pgf/number format/1000 sep = \thinspace},
    nodes near coords align={vertical},
    xtick={0,1},
    legend style={font=\tiny},
    x tick label style={
        rotate=30,anchor=east,/pgf/number format/.cd,
        fixed,
        fixed zerofill,
        precision=2,
        /tikz/.cd
    },
    xmax=1,
    ymax=70,
    xticklabels={yelp, NYC},
    ymin=0,
    enlarge y limits={upper},
    bar width=8pt,
    minor y tick num=5,
    ymajorgrids,yminorgrids,
    ytick align=inside,
    enlarge x limits=0.5,
    every axis plot/.append style={ultra thin},
    legend style={at={(0.5,0.98)},anchor=north},
    legend entries={skewed,original}
]

	\addplot[bar shift=5pt, fill=purple!100, pattern color=white, area legend] table[skip first n=0,x expr={\coordindex},y expr=\thisrowno{1},col sep=comma,each nth point={1},comment chars={T}] {data/skew_skew.csv};
	\addplot[bar shift=5pt, fill=purple!80 ,pattern color=white, area legend] table[skip first n=0,x expr={\coordindex},y expr=\thisrowno{2},col sep=comma,each nth point={1},comment chars={T}] {data/skew_skew.csv};
	\addplot[bar shift=5pt, fill=purple!60,pattern color=white, area legend] table[skip first n=0,x expr={\coordindex},y expr=\thisrowno{3},col sep=comma,each nth point={1},comment chars={T}] {data/skew_skew.csv};
	\addplot[bar shift=5pt, fill=purple!40, pattern color=white, area legend] table[skip first n=0,x expr={\coordindex},y expr=\thisrowno{4},col sep=comma,each nth point={1},comment chars={T}] {data/skew_skew.csv};
	\addplot[bar shift=5pt, fill=purple!20, pattern color=white, area legend] table[skip first n=0,x expr={\coordindex},y expr=\thisrowno{5},col sep=comma,each nth point={1},comment chars={T}] {data/skew_skew.csv};

\resetstackedplots

    \addplot[bar shift=-5pt, fill=darkerblue!100, area legend] table[skip first n=0,x expr={\coordindex},y expr=\thisrowno{1},col sep=comma,each nth point={1},comment chars={T}] {data/skew_orig.csv};
	\addplot[bar shift=-5pt, fill=darkerblue!80, area legend] table[skip first n=0,x expr={\coordindex},y expr=\thisrowno{2},col sep=comma,each nth point={1},comment chars={T}] {data/skew_orig.csv};
	\addplot[bar shift=-5pt, fill=darkerblue!60, area legend] table[skip first n=0,x expr={\coordindex},y expr=\thisrowno{3},col sep=comma,each nth point={1},comment chars={T}] {data/skew_orig.csv};
	\addplot[bar shift=-5pt, fill=darkerblue!40, area legend] table[skip first n=0,x expr={\coordindex},y expr=\thisrowno{4},col sep=comma,each nth point={1},comment chars={T}] {data/skew_orig.csv};
	\addplot[bar shift=-5pt, fill=darkerblue!20, area legend] table[skip first n=0,x expr={\coordindex},y expr=\thisrowno{5},col sep=comma,each nth point={1},comment chars={T}] {data/skew_orig.csv};

    \addlegendimage{black,fill=purple!100, pattern color=white, area legend}
    \addlegendimage{fill=darkerblue, area legend}
\end{axis}
\end{tikzpicture}
}
\vspace{-1.8\baselineskip}
\caption{Time breakdown for different tagging modes (left) and skewed input (right)}
\vspace{-0.8\baselineskip}
\label{fig:tagskew}
\end{figure}

We also analyse the performance of the different tagging modes (see Figure~\ref{fig:tagskew}).
Compared to the original inputs, the skewed inputs in Figure~\ref{fig:tagskew} (right) contain a single record that is $200$ MB in size,
while the remaining records remain the same.
As expected, the use of record-tags (\emph{tagged}) is noticeably slower than the two other tagging modes.
In particular the tagging, partitioning, and type conversion steps take more time,
as they depend on the choice of the tagging mode.
The analysis also highlights the approach's robustness, providing stable performance for the two dissimilar datasets, even if they are skewed (see Figure~\ref{fig:tagskew}).
On the one hand, the time breakdown shows that, except for the type conversion, all steps take roughly the same time for both datasets.
Only type conversion, which involves generating data for more than an order of magnitude more fields in case of the \emph{NYC taxi trips} dataset,
shows perceivable performance differences.
On the other hand, the approach shows robust performance even for highly skewed inputs.

\subsection{End-to-End Parsing}\label{ssec:endtoend}

For the end-to-end parsing experiments, we measured the CPU wall-clock time.
Measurements include the time for reading the input from RAM and writing the parsed data back to system memory.
For \parparaw{}, this includes data transfers between the host and the device.
The end-to-end parsing approach was compared against
\emph{MonetDB},
\emph{Apache Spark},
\emph{pandas},
and the approach presented by M{\"u}hlbauer et al. \cite{2013:instant} (\emph{Inst. Loading}).
In addition, we evaluated the GPU-based parser that is part of NVIDIA's recently introduced open GPU data science project called \emph{RAPIDS}.
For \emph{RAPIDS} we provide two evaluations.
Firstly, simply reading the input into a GPU-based \emph{DataFrame} called \emph{cuDF} from where the data may be queried and processed with \gls{gpu} support (\emph{cuDF*}).
Secondly, exporting the parsed data to the host in the \emph{Apache Arrow} columnar memory format using \emph{cuDF}'s \texttt{to\_arrow()} method (\emph{cuDF}).
\optswitch{MonetDB was built from source, using version 11.31.13 \cite{2005:monetdb}.
To prevent \emph{MonetDB} from storing data on disk, we used a temporary table.
Further, we hinted \emph{MonetDB} at the number of records being parsed to avoid performance degradation due to memory reallocation.}

\begin{figure}[t]
\begin{tikzpicture}
\begin{axis}[
    height=0.6\columnwidth,
    width=\columnwidth,
    ybar=0.5pt,
    xlabel near ticks,
    ylabel near ticks,
    scaled y ticks = false,
    xlabel style={align=center},
    xlabel=partition size (MB),
    ylabel=processing duration (s),
    axis x line*=bottom,
    yticklabel={\pgfmathparse{\tick}\pgfmathprintnumber{\pgfmathresult} },
    y tick label style={/pgf/number format/fixed,
    /pgf/number format/1000 sep = \thinspace},
    nodes near coords align={vertical},
    xtick={0,1,2,3,4,5,6,7,8,9,10},
    x tick label style={
        rotate=90,anchor=east,/pgf/number format/.cd,
        fixed,
        fixed zerofill,
        precision=2,
        /tikz/.cd
    },
    xticklabels={4,8,16,32,64,128,256,512},
    ymin=0,
    enlarge y limits={upper},
    bar width=7pt,
    minor y tick num=5,
    ymajorgrids,yminorgrids,
    ytick align=inside,
    enlarge x limits=0.1,
    legend style={at={(0.98,0.98)},anchor=north east}
]
	\addplot[fill=nyccol, draw=none, area legend] table[skip first n=0, x expr={\coordindex},y expr=\thisrowno{2}/1000,col sep=comma] {data/ete_psize_nyc.csv};
	\addplot[fill=yelpcol, draw=none, area legend] table[skip first n=0, x expr={\coordindex},y expr=\thisrowno{2}/1000,col sep=comma] {data/ete_psize_yelp.csv};
	\legend{NYC taxi trips ($9.1$ GB), yelp reviews ($4.8$ GB)}
\end{axis}
\end{tikzpicture}
\vspace{-0.8\baselineskip}
\caption{End-to-end parsing performance}
\vspace{-0.9\baselineskip}
\label{fig:etepsize}
\end{figure}
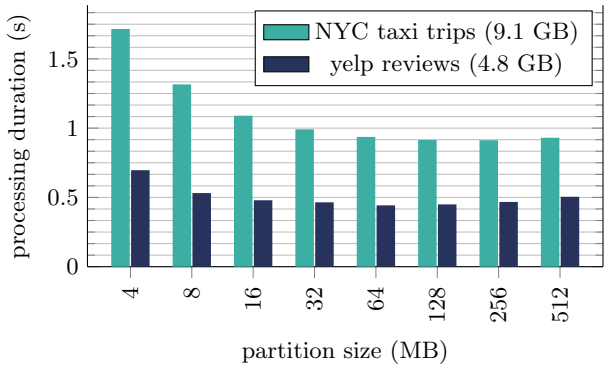

We analyse \parparaw{}'s performance depending on the chosen partition size (see Figure~\ref{fig:etepsize}).
Our evaluation shows that \parparaw{}'s performance increases with the partition size.
Once the partition size grows beyond $128$ MB for the \emph{yelp reviews} and $256$ MB for the \emph{NYC taxi trips} dataset,
however, the end-to-end processing duration starts growing again.
This is due to the increased time for copying the very first partition and returning the parsed data of the very last partition (see Figure~\ref{fig:pipelined}).
Larger datasets compensate the effect of larger partitions.
It is worth noting that this remains the only noticeable effect for larger inputs,
since, with increasing input size, the number of partitions increases linearly, while the time per partition remains the same.

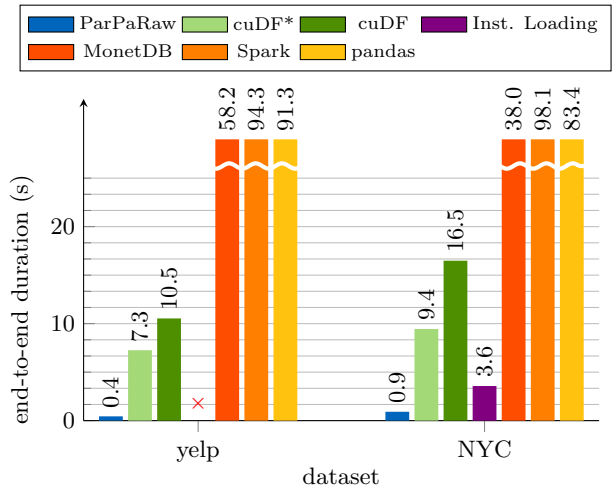
\begin{figure}[t]
\begin{tikzpicture}
\begin{axis}[
    xlabel shift = -4 pt,
    height=0.57\columnwidth,
    width=\columnwidth,
    ybar=2pt,
    xlabel=dataset,
    ylabel=end-to-end duration (s),
    ymax=10,
    axis x line*=bottom,
    axis y line=left,
    yticklabel={
        \pgfmathparse{\tick}\pgfmathprintnumber{\pgfmathresult}
    },
    y tick label style={
        /pgf/number format/fixed,
        /pgf/number format/1000 sep = \thinspace
    },
    restrict y to domain*=0:29, 
    visualization depends on=rawy\as\rawy, 
    nodes near coords,
    every node near coord/.append style={
        rotate=90,
        anchor=west,
        black
    },
    after end axis/.code={ 
        \draw [ultra thick, white, decoration={snake, amplitude=1pt}, decorate] (rel axis cs:0.01,1.05) -- (rel axis cs:0.99,1.05);
    },
    nodes near coords={%
        \pgfmathprintnumber{\rawy}
    },
    clip=false,
    xtick={0,1},
    x tick label style={
    },
    nodes near coords style={/pgf/number format/.cd,fixed, fixed zerofill,precision=1,/tikz/.cd},
    ylabel style={
    },
    xmin=0,
    xmax=1,
    ymax=25,
    y axis line style={shorten >=-30pt},
    xticklabels={yelp, NYC},
    ymin=0,
    bar width=9pt,
    minor y tick num=5,
    ymajorgrids,yminorgrids,
    ytick align=inside,
    enlarge x limits=0.4,
    legend style={at={(0.45,1.44)},anchor=south, legend columns=4,font=\scriptsize},
]
	\addplot[area legend, fill=darkerblue, draw=none] table[x expr={\coordindex},y expr=\thisrowno{1}/1000.0),col sep=comma] {data/ete_all.csv};
    \addlegendentry{ParPaRaw}
    \addplot[area legend, fill=lightergreen, draw=none] table[x expr={\coordindex},y expr=\thisrowno{7}/1000.0),col sep=comma] {data/ete_all.csv};
    \addlegendentry{cuDF*}
    \addplot[area legend, fill=darkergreen, draw=none] table[x expr={\coordindex},y expr=\thisrowno{8}/1000.0),col sep=comma] {data/ete_all.csv};
    \addlegendentry{cuDF}
    \addplot[nodes near coords={\ifnum\coordindex=0 \textbf{\texttt{\small\color{red}$\times$}}\else{\pgfmathprintnumber[/pgf/number format/.cd,fixed, fixed zerofill,precision=1,/tikz/.cd]{\pgfplotspointmeta}}\fi}, area legend, fill=violet, draw=none] table[x expr={\coordindex},y expr=\thisrowno{2}/1000.0),col sep=comma] {data/ete_all.csv};
    \addlegendentry{Inst. Loading}
	\addplot[area legend, fill=caliorange, draw=none] table[x expr={\coordindex},y expr=\thisrowno{3}/1000.0),col sep=comma] {data/ete_all.csv};
    \addlegendentry{MonetDB}
	\addplot[area legend, fill=orange, draw=none] table[x expr={\coordindex},y expr=\thisrowno{4}/1000.0),col sep=comma] {data/ete_all.csv};
    \addlegendentry{Spark}
	\addplot[area legend, fill=nevadagold, draw=none] table[x expr={\coordindex},y expr=\thisrowno{5}/1000.0),col sep=comma] {data/ete_all.csv};
    \addlegendentry{pandas}
\end{axis}
\end{tikzpicture}
\vspace{-0.3\baselineskip}
\caption{End-to-end performance comparison}
\vspace{-0.3\baselineskip}
\label{fig:baselines}
\end{figure}

Figure~\ref{fig:baselines} shows the time taken for parsing the respective input end-to-end.
The performance numbers reported for parsing the $4.8$ GB from the \emph{yelp reviews} dataset highlight the strength of \parparaw{},
which takes only $0.44$ seconds for the more challenging dataset.
Only \emph{cuDF}, which is still roughly $16$ times slower than \parparaw{}, provides comparable performance.
All CPU-based approaches, i.e., \emph{MonetDB}, \emph{Spark}, and \emph{pandas}, are more than two orders of magnitude slower.
Unfortunately, the implementation of \emph{Inst. Loading} provided to us by the authors could not handle the yelp dataset due to its incomplete handling of quoted strings in parallel loads.
Compared to \emph{yelp reviews}, parsing of the \emph{NYC taxi trips} dataset is easier to parallelise, as all line breaks correspond to record delimiters, making it trivial to identify the parsing context.
Hence, even though parsing of the \emph{NYC taxi trips} is computationally more expensive due to its many numerical and temporal fields,
all CPU-based approaches benefit from the simpler format and see great improvements in the parsing rate.
In particular, \emph{Inst. Loading}, the approach proposed by M{\"u}hlbauer et al. \cite{2013:instant}, is about an order of magnitude faster than any other CPU-based implementation.
Even though \emph{Inst. Loading} is able to exploit the parallelism of the $32$ physical cores for the \emph{NYC taxi trips} dataset,
\parparaw{} running on a single \gls{gpu} is still roughly four times faster, despite the fact that \parparaw{} performs a full \gls{dfa} simulation to keep track of the parsing context.
Compared to the remaining approaches, \parparaw{} is more than ten times faster than \emph{RAPIDS} loading the data into \emph{cuDF}
and over $40$ times faster than the next best CPU-based approach.

%% file: chap_conclusions.tex
\section{Conclusions}

\changed{
\begin{sloppypar}
This work presents \parparaw{}, a novel, massively parallel approach to parsing delimiter-separated formats.
Other than the state-of-the-art that targets multicore and distributed systems with a coarse-grained approach, \parparaw{} is designed for fine-grained parallelism.
Supporting parallelism even beyond the granularity of individual records makes it suitable for \glspl{gpu} and ensures load balancing despite small chunks or large and varying record sizes.
Being designed for scalability from the ground up with a data parallel approach that does not require any serial work,
\parparaw{} is future-proof and can continue to gain speed-ups, as more cores are being added with future processors.
Our approach identifies the parsing context (quotation scopes, comments, directives, etc.)
without requiring a prior sequential pass.
\parparaw{} is flexible and generally applicable.
It supports even complex formats with involved parsing rules, as \parparaw{} is able to perform a massively parallel \gls{fsm} simulation.
State-of-the-art \gls{json} parsers and the speculative approach by Ge et al., in contrast, have to deviate from the classic approach of using an \gls{fsm} in order to be able to use \gls{simd} vectorisation and speculation, respectively.
This limits their applicability to other formats and requires designing completely different algorithms when confronted with another format (e.g., log files).
\optswitch{They use bitmap indexes to record the occurrence of structural characters, such as double-quotes, and have to rely on format-specific characteristics.
For instance, they infer the beginning and end of a string enclosed in double-quotes from looking at the odd and even number of set bits, respectively.}
\end{sloppypar}}

We show that \parparaw{} provides scalability without sacrificing applicability and flexibility.
Achieving a parsing rate of as much as $14.2$ GB/s, our experimental evaluation shows that \parparaw{} is able to scale to thousands of cores and beyond.
With \parparaw{}'s end-to-end streaming approach, we are able to exploit the full-duplex capabilities of the PCIe bus while hiding latency from data transfers.
For end-to-end workloads, \parparaw{} parses $4.8$ GB of \emph{yelp reviews} in as little as $0.44$ seconds, including data transfers.
\optswitch{Comparing this to the $0.41$ seconds it would take for transferring the input to the \gls{gpu} alone,
highlights that \parparaw{} successfully maxes out the full-duplex capabilities of the PCIe bus while simultaneously parsing data on the \gls{gpu} with its streaming approach.}
\optswitch{With a total of $0.91$ seconds, similar performance is shown for parsing $9.1$ GB from nyc taxi trips dataset.}


%% file: chap_appendix.tex

%% file: chap_acknowledgements.tex
\section{Acknowledgments}
This research has been supported in part by the Alexander von Humboldt Foundation.
We would like to thank the authors of \emph{"Instant loading for main memory databases"}, in particular Thomas Neumann,
for providing their implementation \cite{2013:instant}.

\balance